\documentclass[traditabstract]{aa}
\usepackage{multirow}
\usepackage{rotating}
\usepackage{graphicx}
\usepackage{amsmath}
\usepackage{natbib}
\usepackage[varg]{txfonts}
\usepackage{psfrag}
\usepackage{units}
\usepackage[colorinlistoftodos]{todonotes}
\def\sun{\hbox{$\odot$}}

\def\3cm{\rm {cm^{-3}}}
\def\2cm{\rm {cm^{-2}}}
\def\s-1{\rm {s^{-1}}}
\def\etal {et\,al.}

\def\kms {\hbox{${\rm km\,s}^{-1}$}}
\def\Kkms {\hbox{${\rm K}\,{\rm km}\,{\rm s}^{-1}$}}

\def\c18o{\hbox{C$^{18}$O}}

\def\ci{\hbox{{\rm [C {\scriptsize I}]}}}
\def\cii{\hbox{{\rm [C {\scriptsize II}]}}}
\def\nii{\hbox{{\rm [N {\scriptsize II}]}}}
\def\hn{\hbox{{\rm H {\scriptsize I}}}}
\def\hii{\hbox{{\rm H {\scriptsize II}}}}

\usepackage{color}

\begin{document}

\title{Carbon gas in SMC low-metallicity star-forming regions}
\author{M.~A.~Requena-Torres\inst{1}\inst{5} \and
        F.~P.~Israel\inst{2} \and
                Y.~Okada\inst{3} \and
                R.~G\"usten\inst{1} \and 
        J.~Stutzki\inst{3} \and
        C.~Risacher\inst{1} \and 
        R.~Simon\inst{3} \and 
        H.~Zinnecker \inst{4}
}
\offprints{M. A. Requena-Torres}
\institute{
 Max-Planck-Institut f\"ur Radioastronomie, Auf dem H\"ugel 69, 53121 Bonn, Germany -
 \email{mrequena@stsci.edu}
\and Sterrewacht Leiden, University of Leiden, Postbus 9513, 2300 RA Leiden, The Netherlands
\and I. Physikalisches Institut der Universit\"at zu K\"oln, Z\"ulpicher Strasse 77, 50937 K\"oln
\and SOFIA Science Center, Deutsches SOFIA Institut, NASA Ames Research Center, Moffett Field, CA 94035, USA
\and Space Telescope Science Institute, 3700 San Martin Dr., Baltimore, 21218 MD, USA}
\date{Received  / Accepted  }
\titlerunning{Velocity-resolved $\cii$ in the SMC}

%
\abstract{This paper presents $\cii$, $\ci$ and CO emission line maps of the star-forming regions N~66, N~25+N~26, and N~88 in the metal-poor Local Group dwarf galaxy SMC. The spatial and velocity structure of the large $\hii$ region N~66 reveals an expanding ring of shocked molecular gas centered on the exciting star cluster NGC~346, whereas a more distant dense molecular cloud is being eroded by UV radiation from the same cluster. In the N~25+N~26 and N88 maps, {\it \textup{diffuse}} $\cii$ emission at a relatively low surface brightness extends well beyond the compact boundaries of the\textup{\textup{ $\it \text{bright}$ }}emission associated with the $\hii$ regions.
In all regions, the distribution of this bright $\cii$ emission and the less prominent $\ci$ emission closely follows the outline of the CO complexes, but the intensity of the $\cii$ and $\ci$ emission is generally anticorrelated, which can be understood by the action of photodissociation and photoionization processes. Notwithstanding the overall similarity of CO and $\cii$ maps, the intensity ratio of these lines varies significantly, mostly due to changes in CO brightness. $\cii$ emission line profiles are up to $50\%$ wider in velocity than corresponding CO profiles. A radiative transfer analysis shows that the $\cii$ line is the dominant tracer of (CO-dark) molecular hydrogen in the SMC. CO emission traces only a  minor fraction of the total amount of gas. The similarity of the spatial distribution and line profile shape, and the dominance of molecular gas associated with $\cii$ rather than CO emission imply that in the low-metallicity environment of the SMC the small amount of dense molecular gas traced by CO is embedded in the much more extended molecular gas traced only by $\cii$ emission.  The contribution from neutral atomic and ionized hydrogen zones is negligible in the star-forming regions observed. The data are available 
at the CDS via anonymous ftp to cdsarc.u-strasbg.fr (130.79.128.5) or via http://cdsweb.u-strasbg.fr/cgi-bin/qcat?J/A+A/}
\maketitle

\section{Introduction}

Clouds of molecular hydrogen (H$_{2}$) gas are host to the first stages of star formation. Their physical conditions are therefore of paramount interest, but, as is well known, direct observation of this gas is practically impossible. Instead,  the properties of the H$_{2}$ gas are derived from observations of tracers, such as the carbon monoxide molecule and its isotopes, or the mixed-in warm dust. Because H$_{2}$ and CO react differently to changes in their environment (e.g., UV radiation, density, and chemistry), it is necessary to fully determine the environment-dependent CO tracer characteristics before more general molecular gas conditions are inferred. In particular, we need to address how well the CO transitions trace H$_{2}$ under varying conditions. This can be done by measuring the emission of $^{12}$CO and its isotopes in various rotational transitions together with that of its dissociation products C$^{\rm o}$ and C$^{+}$. The two neutral carbon $\ci$ lines at 492 and 809 GHz and CO lines with rest frequencies of up to about 900 GHz can be measured from the ground (although observing opportunities rapidly decrease with increasing frequency), but the ionized carbon $\cii$ fine-structure line at 1.9 THz requires an airborne platform such as {\it SOFIA} or one in space, such as {\it Herschel}.

The balance between carbon monoxide and its dissociation product atomic carbon depends on shielding by itself, by molecular hydrogen, and by dust, as well as on the energy density of the impinging radiation field.  With a CO dissociation energy of 10.6 eV and a C$^{\rm o}$ ionization potential of 11.3 eV, the balance between C$^{\rm o}$ and C$^{+}$ in turn also depends closely on irradiation and shielding. For a fixed H$_{2}$ column density, both dust shielding and CO self-shielding inversely scale with ambient metallicity. It is thus of interest to compare similar sets of C$^{+}$, C$^{\rm o}$, and CO measurements spanning a range of metallicity different than Galactic, such as provided by the Large and the Small Magellanic Clouds with $Z=0.5Z\sun$ and $0.2Z\sun$, respectively \citep{russell92}.

A first such study was published by \cite{okada15}, who presented a set of maps of the bright star-forming complex N~159 in the Large Magellanic Cloud (LMC). In this paper, we present similar maps of the bright star-forming complex N~66 in the lower metallicity Small Magellanic Cloud (SMC), as well as strip-maps for the less active star-forming regions N~25+N~26 and N~88, also in the SMC. N~66 is the largest and brightest $\hii$ region complex, somewhat isolated in the northern part of the SMC Bar. At any wavelength, it is much less bright than its counterpart 30 Doradus in the LMC, but it is similar to the above-mentioned N~159. N~25, N~26, and N~88 are relatively compact HII regions in the crowded southwest of the SMC Bar, and in the thinly populated SMC Wing, respectively. 
\begin{figure}[!hpt]
\begin{center}
\includegraphics[angle=0,width=7.5cm]{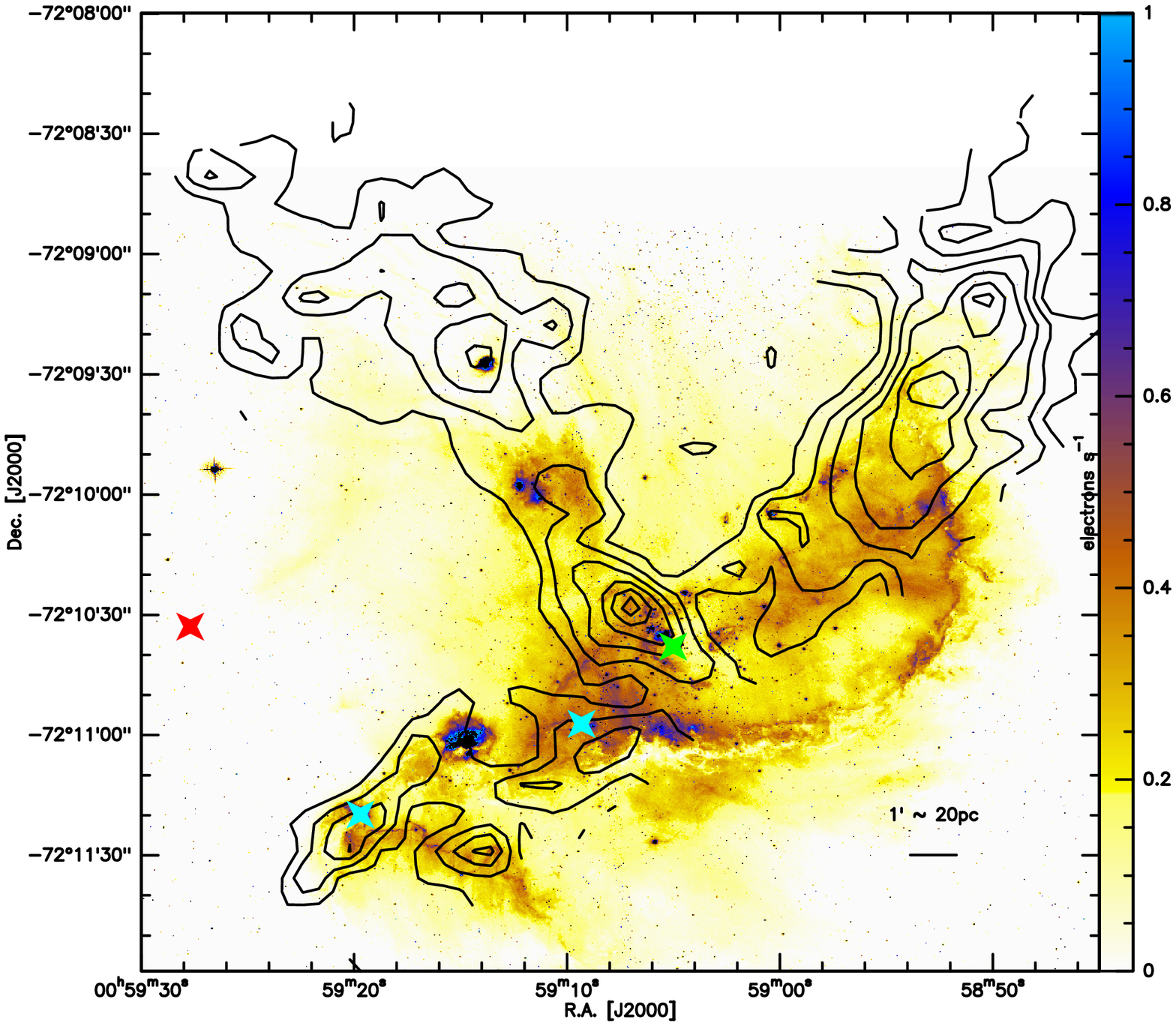}
  \caption{\footnotesize{H$\alpha$ map in N66 observed by the Hubble Space Telescope \citep[ACS WFC1 archival data, ][]{nota06}. The units are electrons per second, and the conversion factor $1.99\times10^{-18}$ $ergs~ cm^{-2} Ang^{-1} electron^{-1}$. We show  superposed in black the contours of the $\cii$ integrated intensity (in K \kms, integrated over the LSR velocity interval of 141 to 166 \kms) from the present work. We have marked the position of the stellar cluster NGC346 by a green star, detected water masers by blue stars, and the supernovae by red star. }}
  \label{HAlpha}
\end{center}
\end{figure}

N~66 is a very extended (100 - 150 pc) diffuse and filamentary gas cloud with a bright core of about 50 pc (see Fig. \ref{HAlpha} for a comparison of the new [CII] data with the archival HST H$\alpha$ emission). The core consists of a long ionization front in the form of a curved ridge with various bright patches of ionized gas scattered about, centered on the prominent star cluster NGC~346, which is the main source of ionization. At the southeastern edge of the ridge is a bright patch known as N66~A by \cite{heydari10}; the bright area at the opposite, northwestern edge of the ridge is called N66~B, and there is another, slightly detached patch to the northeast of the ridge, which is referred to as N66~C, a nomenclature we follow here \citep[see Fig. 2 in ][]{reid06}. The stellar content of N~66 and NGC~346 has been determined in great detail; almost 100 000 stars have been identified on HST images, including a large number of pre-main-sequence stars \citep[][]{nota06,Gouliermis06}. N~66 is quite evolved because star formation has been taking place here for 10-20 million years \citep[][]{sabbi07,cignoni11,demarchi11}, even though the cluster NGC~346 that contains at least a dozen O stars \citep{evans06} is at most three million years old. The large pre-main-sequence population, the two H$_{2}$O masers detected in N~66 \citep{breen13}, at least one supernova remnant \citep{ye91}, and the large number of embedded young stellar objects \citep{simon07} all suggest that N~66 is still forming stars, perhaps triggered by stars of an earlier generation \citep{gouliermis08}, but at lower rates than in the past \citep{cignoni11}.

   \begin{table*}
      \caption[]{Observation summary}
         \label{tableObsSum}
                \begin{center}
        \begin{tabular}{llllllll|l|}
            \noalign{\smallskip}
            \hline
            \noalign{\smallskip}
            Species      &  Transition &Freq. [GHz]  & instrument  & $\eta_f$ & $\eta_{mb}^1$&HPBW [$''$] \\
            \noalign{\smallskip}
            \hline
            \noalign{\smallskip}
                CO      & J=2-1 & 230.538 & SHeFI  &0.95 & 0.75& 27.3 \\
                        & J=3-2 & 345.795 & FLASH+  &0.95 & 0.67(0.69) & 18.2 \\ 
                        & J=4-3 & 461.040 & FLASH+  &0.95 & 0.58(0.63) & 13.6\\ 
                        & J=6-5 & 691.473 & CHAMP+  &0.95 & 0.56 & 9.1\\
        $^{13}$CO    & J=3-2 & 330.587 & FLASH+ &0.95 & 0.67(0.69) & 19.0\\ 
        C$^{18}$O$^2$    & J=3-2 & 329.331 & FLASH+ &0.95 & 0.67(0.69) & 19.0\\ 

            \hline
            \noalign{\smallskip}
        $\ci$   & $^3$P$_1$ - $^3$P$_0$ & 492.160& FLASH+ &0.95 & 0.58 (0.63) & 12.8 \\ 
                $^2$    & $^3$P$_2$ - $^3$P$_1$ & 809.341& CHAMP+ &0.95 & 0.47 & 7.8 \\ 
            \hline
            \noalign{\smallskip}
        $\cii$  & $^2$P$_{3/2}$ - $^2$P$_{1/2}$ & 1900.536      &GREAT &0.97 & 0.67 & 14.1\\ 
    $\nii^3$  & $^3$P$_1$ - $^3$P$_0$ & 1462.145  &GREAT &0.97 & 0.67 & 18.3\\ 
            \noalign{\smallskip}
            \hline
            \end{tabular}
                \end{center}
Note: $^1$ Values from 2013, in () from 2014. $^2$ Very limited observation and/or undetected. $^3$ Only observed in N~25+N~26 and N88, undetected.\\
Technical data from:\\
http://www3.mpifr-bonn.mpg.de/div/submmtech/heterodyne/flashplus/flashmain.html\\http://www3.mpifr-bonn.mpg.de/div/submmtech/heterodyne/champplus/champ\_efficiencies.16-09-14.html\\http://www3.mpifr-bonn.mpg.de/div/submmtech/heterodyne/great/GREAT\_calibration.html
      \end{table*}

Particularly relevant to the purpose of this paper are the previous studies of the N~66 interstellar medium (ISM). The $\hii$ region-cum-SNR was mapped in the radio continuum and the H$\alpha$ line \citep{ye91,danforth03,reid06}.  Mid-infrared images, photometry, and spectroscopy were published by \citet[ ISO]{contursi00} and \citet[ Spitzer]{whelan13} concentrating on PAH behavior; inferred UV radiation field strengths are up to 10$^{5}$ times greater than in the solar neighborhood. Finally, maps of cool low-J CO and hot H$_{2}$ were presented by \cite{rubio00}, whereas the $\cii$ emission in the 158$\mu$m line was mapped at low spectral resolution by \cite{israel11}. The most remarkable result from a comparison of the H$_{2}$, PAH, CO, and $\cii$ maps with the distribution of pre-main-sequence stars as depicted by  \cite{hennekemper08} or \cite{gouliermis14}, for instance, is their strong overall resemblance, even though there are some differences in detail.

N~25 is one of a group of $\hii$ regions also containing the compact N~26, the large and diffuse N~22, and N~21 and N~23 \citep{testor01,testor14} in the crowded southwestern SMC Bar. The ionization of each of the nebulae appears to be dominated by a single O star. The complex and its surroundings were mapped in CO by \citet[SMC-B2 no.3]{rubio93}, in the far-infrared continuum by \citet[SMCB2 N]{bot10}, and in $\cii$ by \cite{israel11}. N~25+N~26 and N~22 are relatively pronounced radio sources at 843 MHz \citep{mills84}.

%
\begin{figure}[!hpt]
\begin{center}
\includegraphics[angle=0,width=7.5cm]{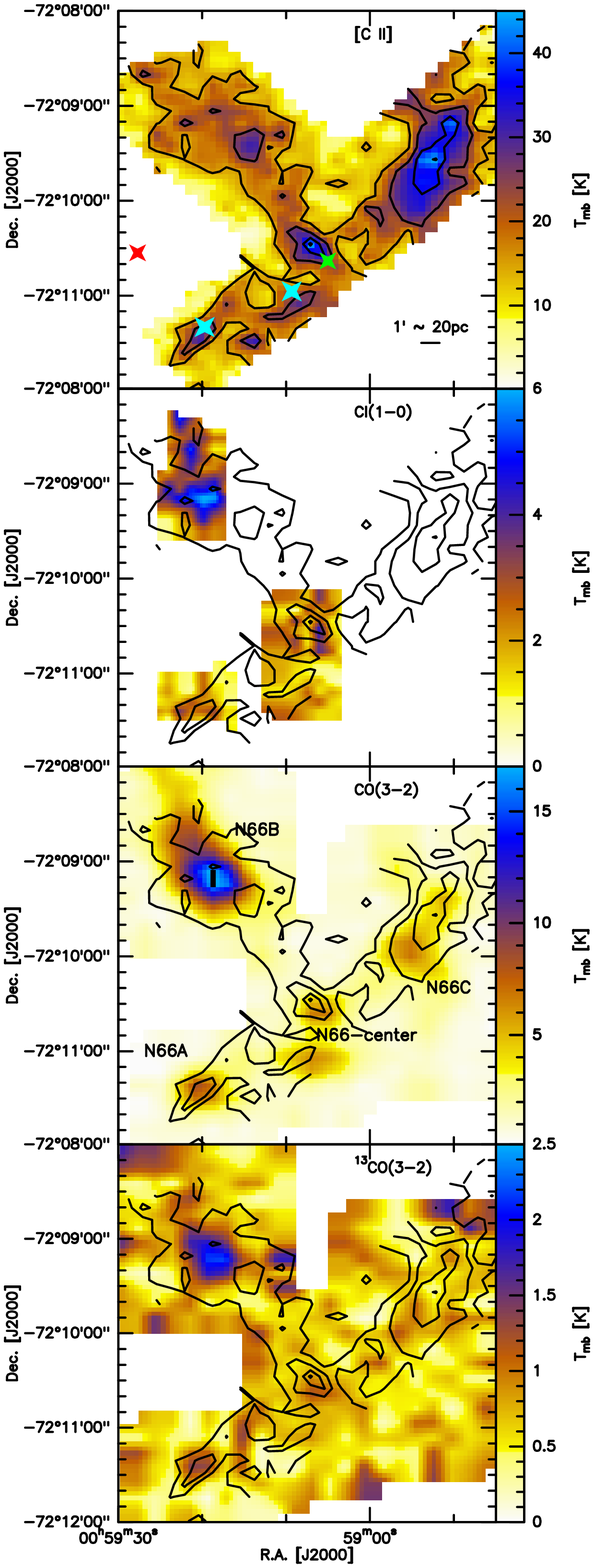}
  \caption{\footnotesize{Integrated intensity maps between LSR velocities of 141 to 166 \kms (in color) of N~66 for the $\cii$, $\ci$ ($^3$P$_1$ - $^3$P$_0$), CO (3-2) and $^{13}$CO (3-2) emission. For comparison, we have superposed the contours of the $\cii$ integrated intensity in black. As in Fig. \ref{HAlpha}, we have marked the star cluster NGC346 by a green star, water masers by blue stars, and the supernova by a red star in the top panel. In the CO(3-2) panel we have marked the different components of the N66 structure discussed in the text.}}
  \label{initmaps1}
\end{center}
\end{figure}

%
\begin{figure}[!hpt]
\begin{center}
\includegraphics[angle=0,width=7.2cm]{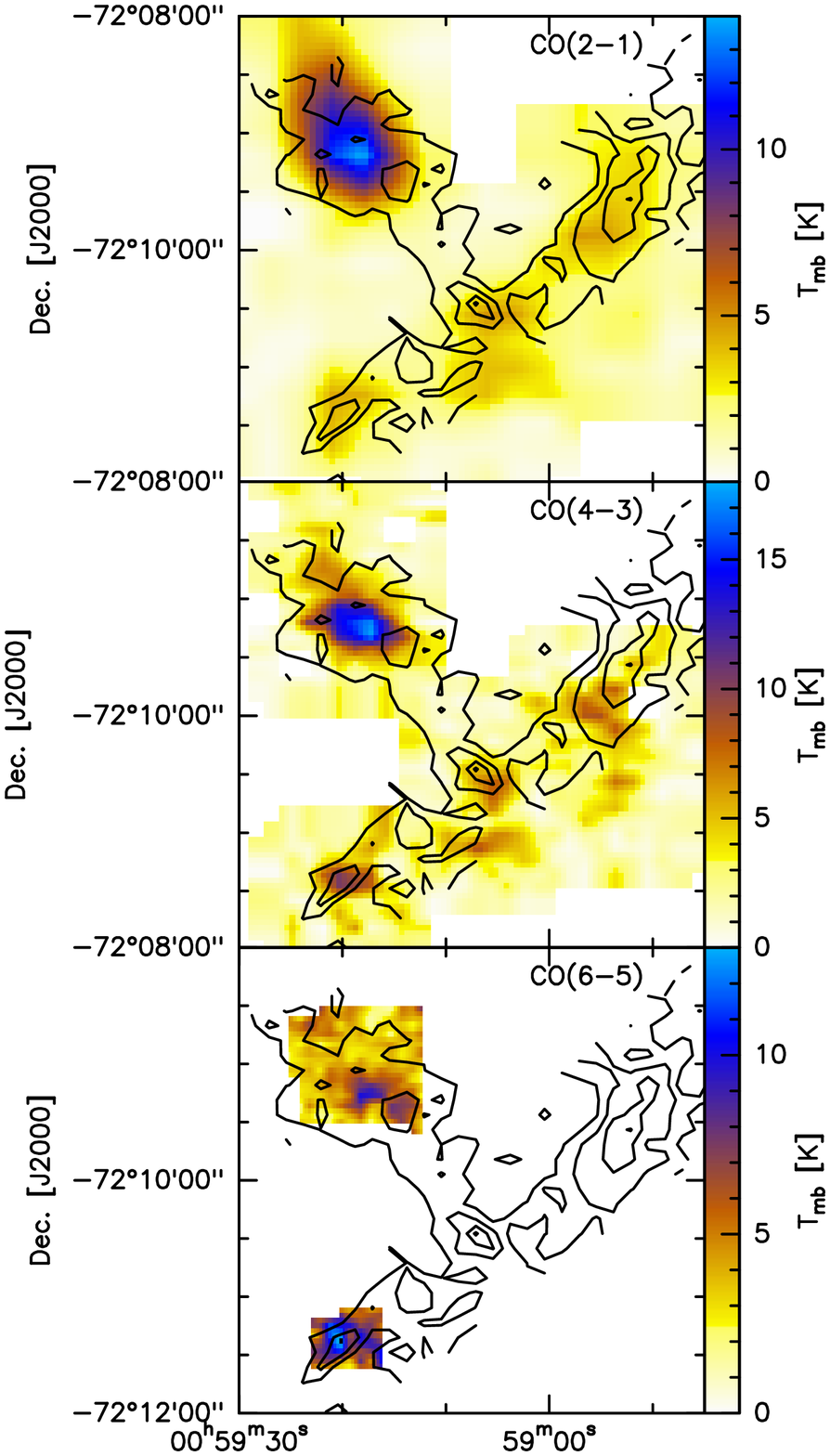}
  \caption{\footnotesize{Integrated intensity maps of N~66 in the CO J = (2-1), (4-3), and (6-5) transitions. Contours as in Fig. \ref{initmaps1}.}}
  \label{SOFIAmaps2}
\end{center}
\end{figure}

In contrast, N~88 is a well-studied rather isolated compact object in the SMC Wing. It is included in many SMC surveys because of its brightness and its isolated nature. It is unusually rich in dust, apparently lacking in small grains, but rich in carbon, suggesting recent photoevaporation of dust particles \citep{heydari99,kurt99}. It is a very young star-forming region, comparable to the Orion nebula in the Milky Way \citep[cf. ][]{testor10}. It has been mapped in CO by \cite{israel03}, and in $\cii$ by \cite{israel11}. 

\section{Observations}

The observations were executed in two different projects, one covering strips in the N~25 and N~88 regions, and one mapping the extended emission in N~66. For the observations of the CO and $\ci$ lines we used the instrumentation on the APEX telescope\footnote{This publication is based in part on data acquired with the Atacama Pathfinder Experiment (APEX). APEX is a collaboration between the Max-Planck-Institut f\"ur Radioastronomie, the European Southern Observatory, and the Onsala Space Observatory} \citep{guesten06}, and for the [CII] and [NII] observations we used the GREAT\footnote{GREAT is a development by the MPI f\"ur Radioastronomie and the KOSMA/ Universit\"at zu K\"oln, in cooperation with the MPI f\"ur Sonnensystemforschung and the DLR Institut f\"ur Planetenforschung.} instrument \citep{heyminck12} onboard SOFIA\footnote{This work is based in part on observations made with the NASA/DLR Stratospheric Observatory for Infrared Astronomy (SOFIA). SOFIA is jointly operated by the Universities Space Research Association, Inc. (USRA), under NASA contract NAS2-97001, and the Deutsches SOFIA Institut (DSI) under DLR contract 50 OK 0901 to the University of Stuttgart} \citep{young12}. Most of the observations were executed in different APEX and SOFIA runs during 2013, but some additional data were observed for N~25 and N~88 in 2014. 

\subsection{SOFIA observations}

%
\begin{figure*}[!hpt]
\begin{center}
\includegraphics[angle=0,width=15cm]{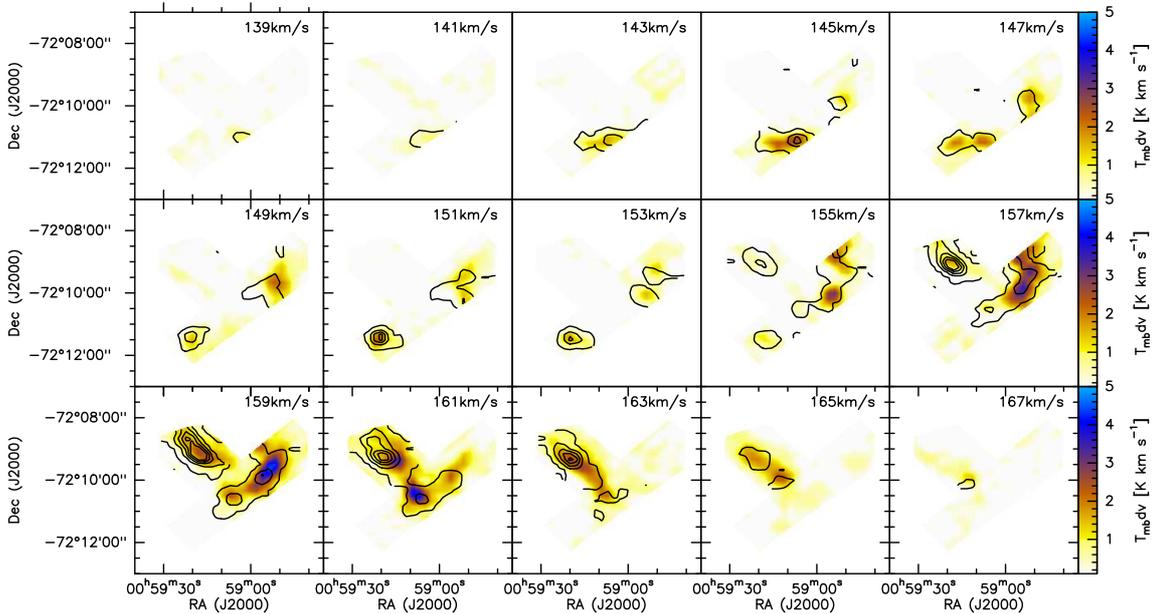}
\caption{\footnotesize{N~66 $\cii$ (color) and CO(3-2) channel maps (contours from 0.2 $\Kkms$ in steps of 0.4 $\Kkms$) beam-averaged to a $20''$ grid. Maps were velocity-smoothed to 2 $\kms$. The panels show the integrated emission in the velocity range of 139 to 167 $\kms$ in intervals of 2 $\kms$. Note the general resemblance of the two distributions, even though there are some
differences in relative intensity.}}
  \label{specmaps}
\end{center}
\end{figure*}

During the Cycle 1 SOFIA Southern Deployment to New Zealand in July 2013, we used the GREAT receiver to study regions in the SMC. We made strip maps of N~25 and N~88 in on-the-fly (OTF) mode in the $\cii$ line at 1900.5369 GHz (158 $\mu$m) and the $\nii$ line at 1462.145 (205 micron). The  strip maps consist of three parallel strips separated by $8''$ and several arcmin in length; the effective width of the maps is thus $0.5'$. The N~88 strip map is oriented in an east-west direction (PA=90$^{\circ}$) and covers the northern half of the $\hii$ region; the southern edge of the map cuts through the center of the $\hii$ region. The N~25 map is oriented diagonally in a northeast-southwest direction (PA=45$^{\circ}$). It cuts across the main body of the $\hii$ region N~25.

For N~66 we mapped in OTF mode with only one of the GREAT channels (L2 to observe the $\cii$ line) that was connected to a digital XFFTS spectrometer \citep{klein12}, providing a bandwidth of 2.5 GHz. For the mapping, we divided the region into three subregions, one to cover the central region and two side maps to continue along the bar in the north and in the south. The submaps were made with respect to the reference position R.A. 0$^h$ 59$^m$ 20.0$^s$ and Dec. -72$^o$09$'$00.0$''$ (Eq. J2000.0). The middle one centered at $(-38.3'',-31.7'')$ with a size of $120''\times210''$, the southern at $(-13,7'',-151.6'')$ and the northern at $(-154.3'',+4.4'')$, both with size $90''\times84''$.

The nominal focus position was updated regularly against temperature drifts of the telescope structure. The pointing was established with the optical guide cameras to an accuracy of better than a few arcsecs. The beam widths and efficiencies are indicated in Table\,\ref{tableObsSum}. The data were calibrated with the  standard GREAT pipeline \citep{guan12}, removing residual telluric lines, and further processed with the GILDAS package\footnote{http://www.iram.fr/IRAMFR/GILDAS}. This mostly consisted of linear baseline removal, spectral resampling, and gridding to the final data cubes.

\subsection{Complementary APEX observations}

Both N~25 and N~88 were observed with the APEX telescope in May, June, October, and November 2014. The N~66 molecular cloud was mapped in May and June 2013, with the SHeFI \citep{vassilev08}, FLASH+ \citep{klein14} and The Carbon Heterodyne Array of the MPIfR \citep[CHAMP$^+$;][]{Kasemann06} receivers. A summary of the observations is shown in Table \ref{tableObsSum} together with all GREAT/SOFIA observations. With SHeFI we used the APEX1 instrument to map the CO(2-1) transition in OTF mode and to find emission peaks for further observation in the $^{13}$CO and the C$^{18}$O $J$=3-2 transitions (undetected). With FLASH+ we obtained OTF maps in the CO (3-2), $^{13}$CO (3-2), and CO (4-3) lines. For $\ci$ ($^3$P$_1$ - $^3$P$_0$) we observed only the main north region and two other small regions in the Bar, where the emission is weaker. With CHAMP+ we performed limited observations in the CO (6-5) and the undetected $\ci$ ($^3$P$_2$ - $^3$P$_1$) lines. The maps of N66 are centered on R.A. 0$^h$ 59$^m$ 20.0$^s$ and Dec. -72$^o$09$'$00.0$''$ (eq. J2000.0), and the reference offset position free from emission is located at (-240$''$,+200$''$). The maps are presented at full spatial resolution in Figs. \ref{initmaps1} and \ref{SOFIAmaps2}.

For both sets of observations, the focus was checked at the beginning of each observation and pointing was checked every hour to be better than 2$''$. The data from APEX were calibrated to T$^*_a$ units and further reduction such as the conversion to the T$_{mb}$ scale, baseline subtraction, and extraction of the specific slices, spectra or integrated intensities was made with the GILDAS package.

\section{Results and analysis}

\subsection{Large-scale mapping of N~66}

%
\begin{figure*}[!hpt]
\begin{center}
\includegraphics[angle=0,width=15cm]{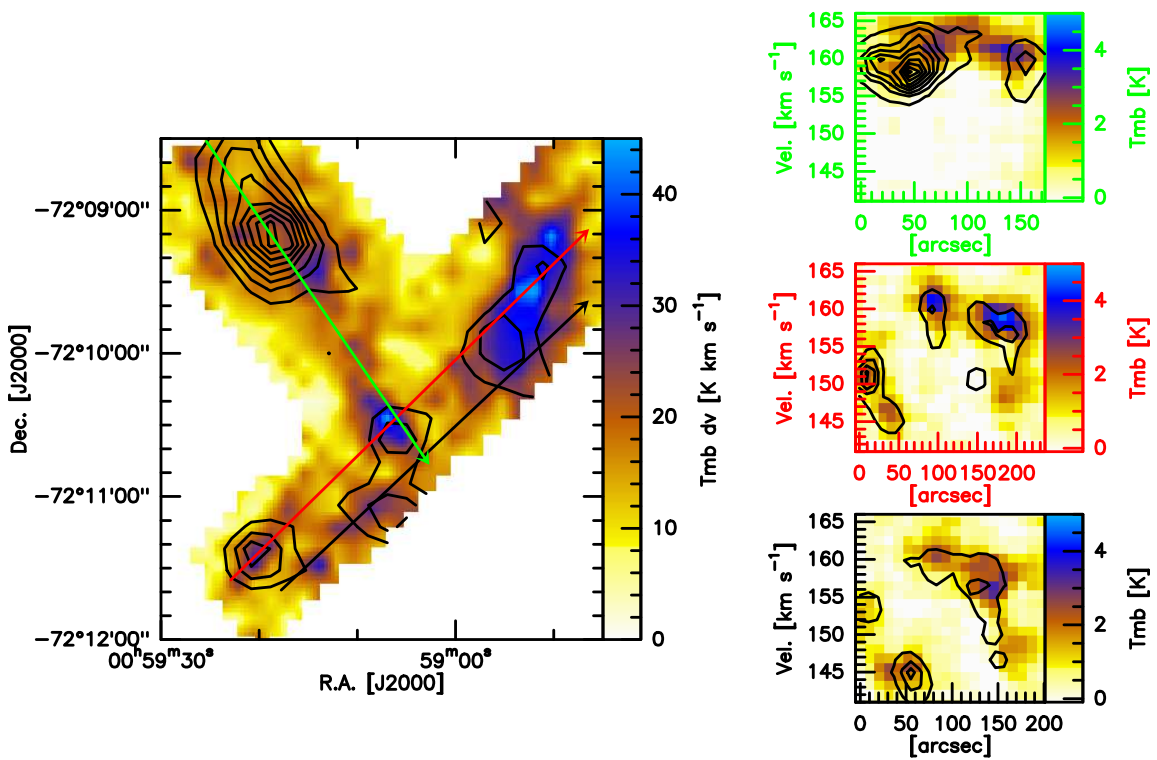}
  \caption{\footnotesize{Diagrams of the velocity structure of the N~66 gas. The left panel shows the velocity-integrated intensity of $\cii$ in color (K km s$^{-1}$ scale), with the integrated intensity of the CO (3-2) transition superposed in black contours (starting at 3 K km s$^{-1}$ increasing in steps of 2 K km s$^{-1}$). The three panels at the right contain the position-velocity maps along the cuts indicated in the left panel, with position (in arcsec) increasing along the direction indicated by the arrows in that panel. Note the ring-like distribution of clumpy emission in the ridge depicted in the lower two panels. The panel at the
top contains the orthogonal cut through the plume, showing more continuous and brighter emission.
}}
  \label{fig:N66slice}
\end{center}
\end{figure*}

\begin{figure*}[!hpt]
\begin{center}
\includegraphics[angle=0,width=17cm]{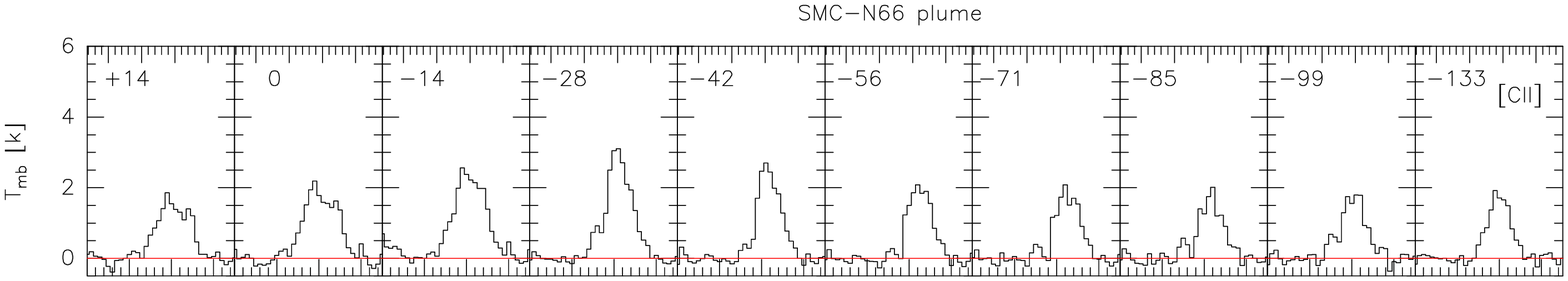}
\includegraphics[angle=0,width=17cm]{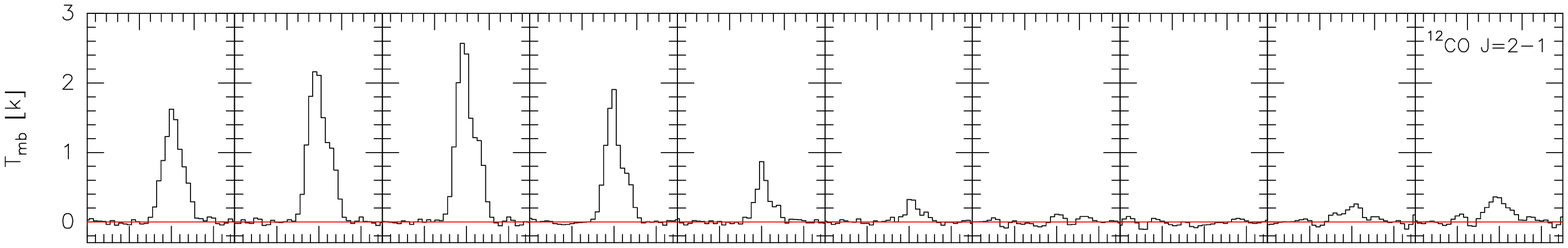}
\includegraphics[angle=0,width=17cm]{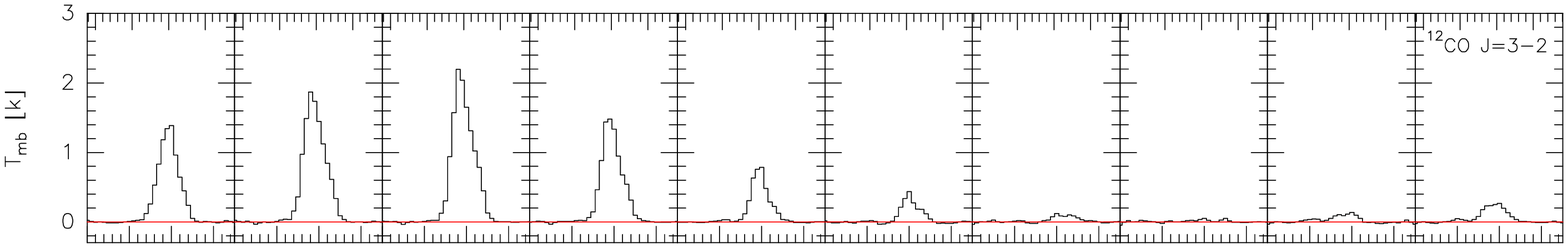}
\includegraphics[angle=0,width=17cm]{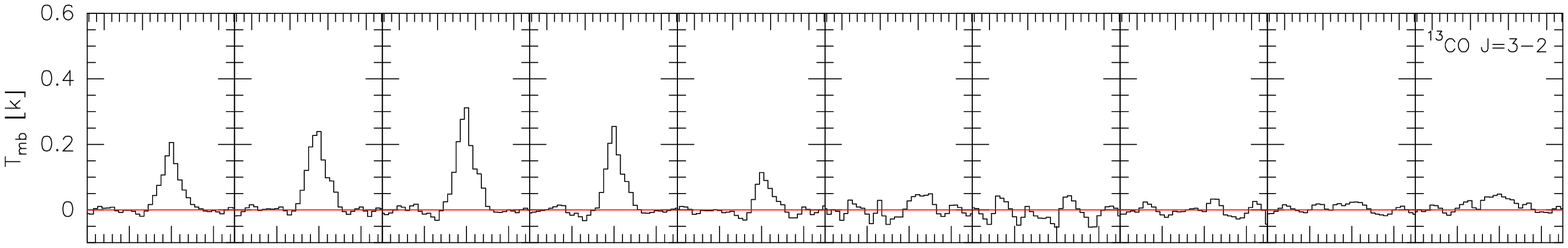}
\includegraphics[angle=0,width=17cm]{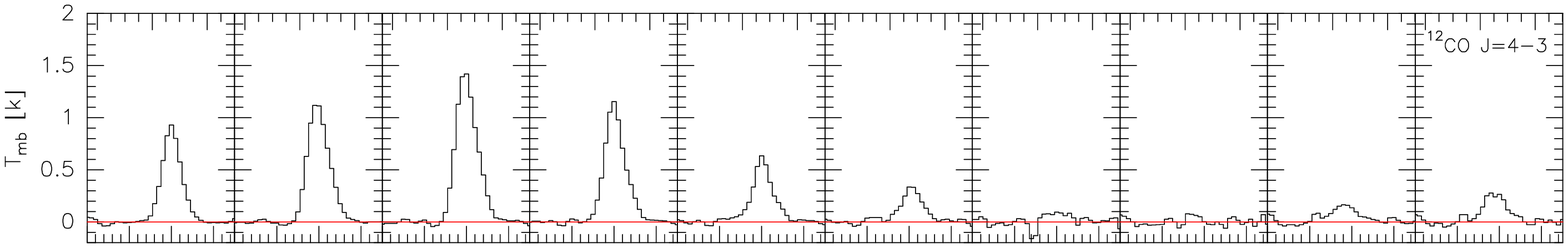}
\includegraphics[angle=0,width=17cm]{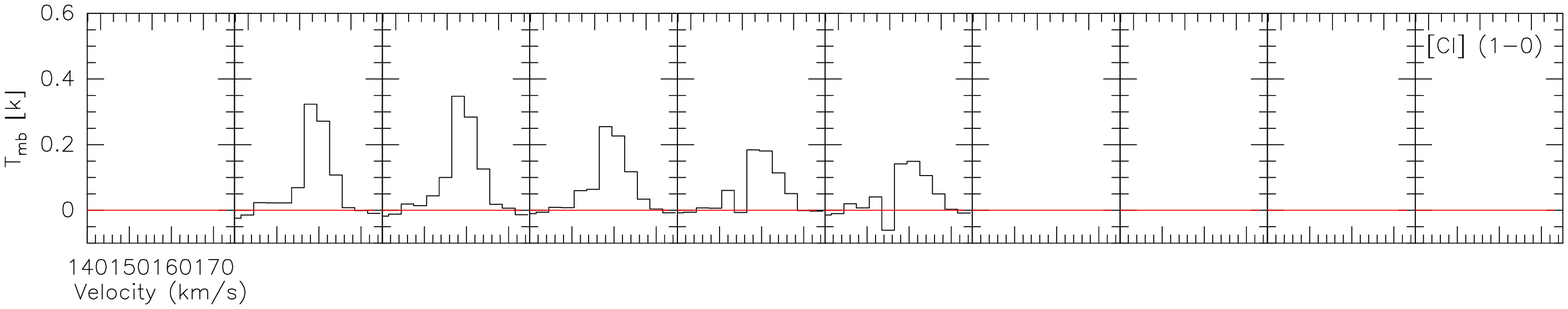}
  \caption{\footnotesize{Line profiles along the N~66 plume, taken from the maps in Figs. \ref{initmaps1} and \ref{SOFIAmaps2}, following the plume emission as shown in Fig. A.3.  
All profiles are at a uniform resolution of $27.3''$. Vertical scales are main-beam brightness temperature $T_{mb}$. Velocities range from $V_{LSR}=140$ km/s to $V_{LSR}=175$ km/s and are marked at intervals of 2 km/s }}
\label{fig:N66pluapp}
\end{center}
\end{figure*}

\begin{figure*}[!hpt]
\begin{center}
\includegraphics[angle=0,width=17cm]{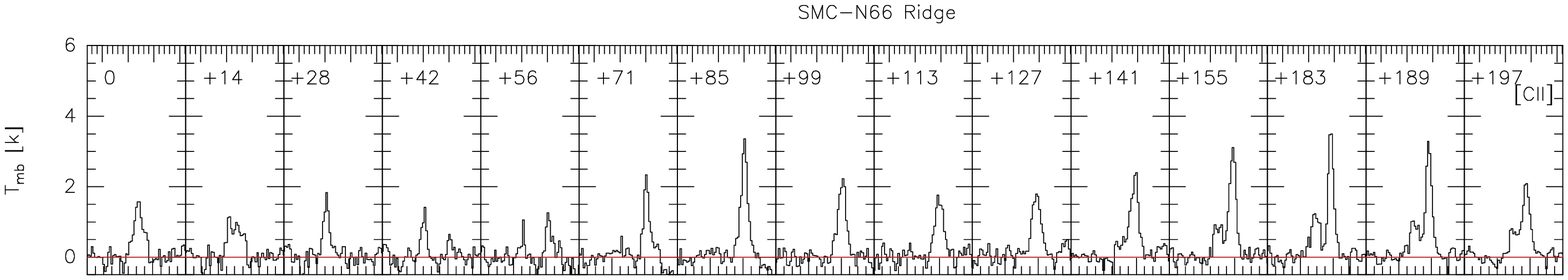}
\includegraphics[angle=0,width=17cm]{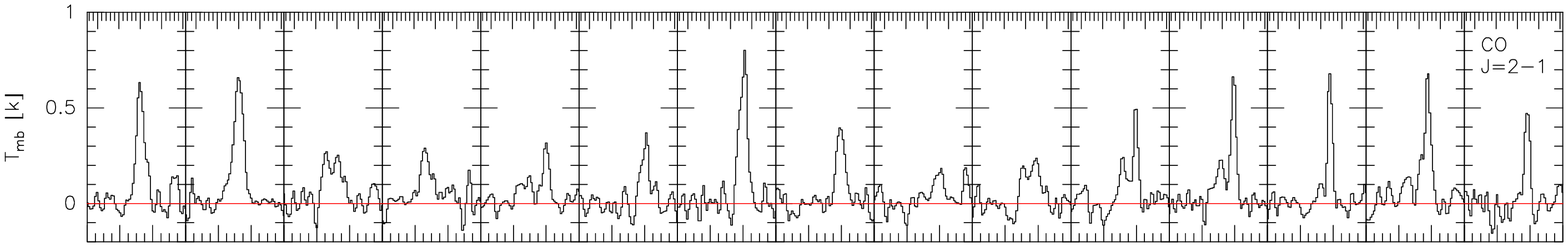}
\includegraphics[angle=0,width=17cm]{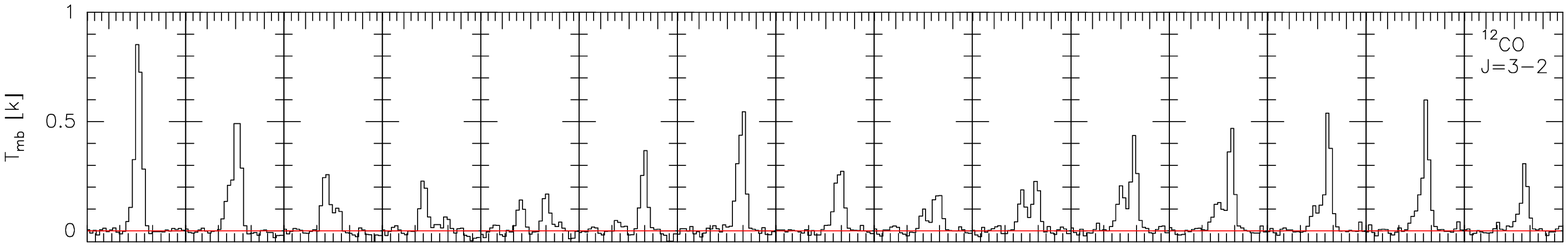}
\includegraphics[angle=0,width=17cm]{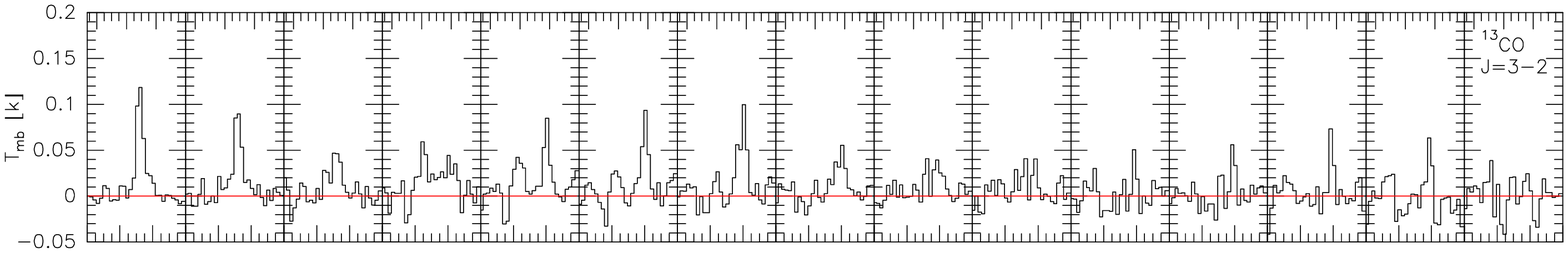}
\includegraphics[angle=0,width=17cm]{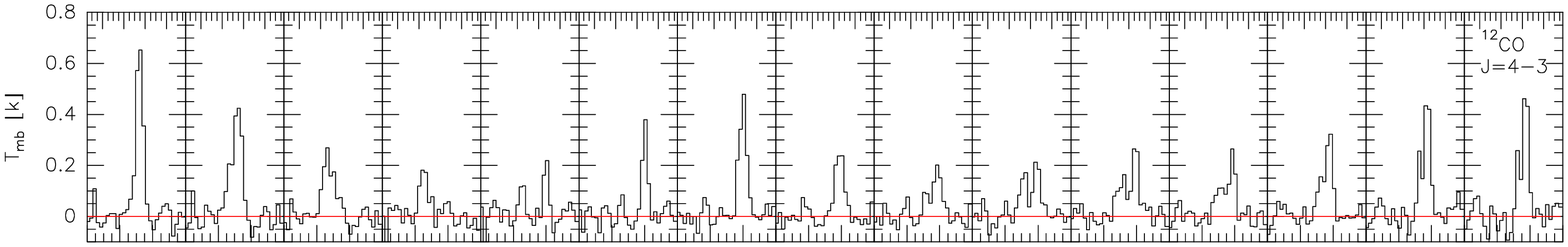}
\includegraphics[angle=0,width=17cm]{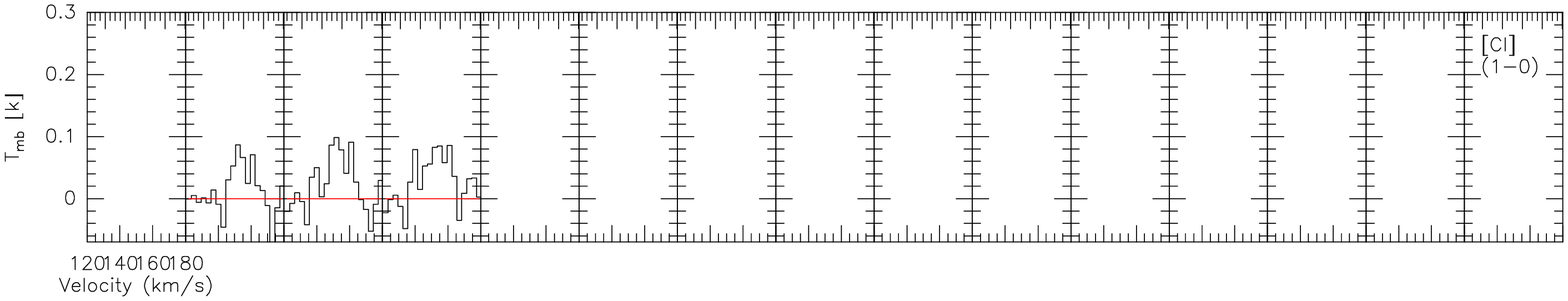}
 \caption{\footnotesize{Line profiles along the N~66 ridge, taken from the maps in Figs. \ref{initmaps1} and \ref{SOFIAmaps2}, following the ridge emission as shown in Fig. A.3. All profiles are at a uniform resolution of $27.3''$. Vertical scales are main-beam brightness temperature $T_{mb}$. Velocities range from $V_{LSR}=120$ km/s to $V_{LSR}=180$ km/s and are marked at intervals of 2 km/s}}
\label{fig:N66ridapp}
\end{center}
\end{figure*}

For the relatively large N~66 complex, we obtained the maps shown in Figs.\,\ref{initmaps1} and \ref{SOFIAmaps2} in the various transitions of ionized carbon, neutral carbon, and carbon monoxide listed in Table\,\ref{tableObsSum}. Because of time constraints combined with the general weakness of the line emission from the N~66 complex, we were unable to obtain complete maps of the whole area. Instead, we concentrated on mapping the two main distinguishing structures in the center of the object: the ionization ridge running from the southeast to the northwest, and the plume-like feature perpendicular to the ridge, running in a northeastern direction. The maps cover the bright central part of the N~66 nebular complex. More diffuse, ionized gas extends far beyond the map boundaries, but most of this appears to be excited by the same stellar clusters that cause the ionization fronts in the map to light up. We detect $\cii$ emission essentially everywhere in the region mapped. This is in contrast to CO, which is present in the form of clumpy emission, embedded in much weaker diffuse emission. Four distinct features can be identified. These are (i) the bright CO cloud at the northern end of the plume (N~66B), (ii) the bright $\cii$ cloud (weak in CO) at the northwestern end of the ridge (N~66C), (iii) a cloud of modest CO and $\cii$ intensities at the southeastern end of the ridge (N~66A), and (iv) a similar cloud at the center of the ridge, where the plume emanates (N~66-center, see the labels in the CO (3-2) map in Fig. \ref{initmaps1}). The {\it APEX} CO (6-5)  and $\ci$ ($^3$P$_1$ - $^3$P$_0$) maps sample only a few fields. Nevertheless, it is very clear that $\ci$ is relatively strong in the plume/N~66B and weak throughout the ridge. With this behavior, $\ci$ follows that of CO rather than that of $\cii$. 

The general distribution of the emission in both CO and $\cii$ is similar to that depicted by \cite{israel11}, but our new maps have higher spatial and spectral resolutions and sensitivity and consequently show better detail\footnote{The central coordinates of the $\cii$ map shown by \cite{israel11} are incorrect. The actual coordinates of the (0,0) position are RA=$00^{h}59^{m}16^{s}$, Dec=$-72^{\circ}25'40''$ (B1950.0)}. Notwithstanding the significant differences in resolution and coverage of N~66, we find that the area-integrated $\cii$ fluxes are roughly similar, with the combined emission of the ridge and the plume in the present map about half of the total emission in the much larger ($5'\times6'$) KAO map. $\cii$ emission is strongest in the ridge, especially in the northwest (N~66C), but the plume is also clearly delineated in $\cii$. In CO, by far the strongest emission in the map is seen at the tip of the plume. The ridge is only weakly seen in CO. This agrees with the {\it SEST} CO (2-1) maps published by \citet{rubio93}, who additionally found a clear correlation between long and narrow fronts of (near-infrared emission from) warm and presumably shocked H$_{2}$ gas and the weak CO ridge. In contrast, the H$_{2}$ emission is very weak toward the plume where CO is strong. 

The $\cii$/CO line intensity ratio changes greatly across the maps. In the ridge this ratio increases from about 2.5 in the southeast to about 5 in the center to about 9 in the $\cii$-bright cloud in the northwest. Similar high $\cii$/CO ratios occur in the bottom of the plume, dropping steeply to values below 2 at its tip. These high ratios are entirely caused by the lack of CO flux, as the $\cii$ flux hardly changes across the entire plume region (Fig.\,\ref{initmaps1}).  The $\ci$ emission is very weak toward the ridge, but becomes conspicuous toward N-66C, coincident with the strong CO emission.

In Fig. \ref{fig:N66slice}  we present slices through the data cube, showing the emission distribution in the velocity domain. We show two parallel cuts through the ridge and a single cut through the plume. In these figures the color image represents the $\cii$ emission, whereas CO and $\ci$ emission are indicated by black and white contours, respectively. We note that the $\ci$ was not mapped in the northwestern half of the ridge. The plume is restricted to a single, relatively narrow velocity component and more or less continuous in spatial position. However, the ridge shows three distinct velocity components, at 145 km s$^{-1}$, at 150 km s$^{-1}$, and at 160 km s$^{-1}$ with a clumpy distribution in space.

Especially in the southernmost of the two cuts, the $\cii$ emission may be somewhat displaced from the CO emission. More generally speaking, the occurrence of emission in the velocity domain is the same for the various lines, except for some weak $\cii$ at about 150 km s$^{-1}$ that seems to have no CO counterpart.

In Fig. \ref{fig:N66pluapp} and \ref{fig:N66ridapp} we show spectra for cuts in two orthogonal directions across the axis of the plume and of the ridge components for N 66 (for the orientation of the cuts, see Fig. A.3 in the appendix). These profiles show quite close similarities in most of the positions of the cuts.

\subsection{Carbon column densities in N~66}

\subsubsection{Results from LTE modeling}

In the absence of information constraining the effect of non-equilibrium processes in the interstellar medium studied, quantitative parameters can be estimated using the local thermodynamic equilibrium (LTE) approximation that ignores the presence of such processes. In the LTE approximation, the optical depth of emission from a species varies as a function of its column density and its excitation temperature; the emission varies as a function of optical depth and excitation temperature. In a previous paper, \cite{okada15} have presented a procedure to produce maps of estimated column densities derived from the observed line emission cubes of the LMC star-forming complex N~159. Here, we apply the procedure outlined in this paper to our results on N~66. To derive carbon column densities and the column density of H$_2$ implied by these, we start by assuming that LTE conditions apply for each species (C$^{+}$, C$^\circ$, $^{13}$CO, and $^{12}$CO) and by assuming that the excitation temperature T$_{ex}$ is constant over each map-pixel. We assumed different beam filling factors ($\eta$=0.1, 0.2, 0.5 and 1) to try to understand the significance of a varying beam filling on the results.  

We used the observed CO (3-2) transitions to calculate the contribution of CO to the total C. As the  $^{13}$CO isotopolog has a low optical depth, we can use this to derive, assuming that both isotopologues have the same $T_{ex}$ and $\eta$:

\begin{equation}
T_{ex} = \frac{h \nu}{k} \left[ ln \left\{ \left( 1- e^{-\tau_{13}} \right) \eta  \frac{h \nu}{k T_{mb}(^{13}CO)}+1 \right\} \right]^{-1}
.\end{equation}

The actual optical depth  $\tau_{13}$ is derived from the observed isotopolog peak temperature ratio $T_{B}(CO(3-2))/T_B(^{13}CO(3-2))$ equaling $\left[ 1-\exp \left( - \tau_{12} \right) \right] / \left[ 1-\exp \left( - \tau_{13} \right) \right]$. As in \cite{okada15}, we used the isotope ratio (see Sect. 3.2.2) to reduce the parameter space with $\tau_{12} = 50 \tau_{13}$.

The column density in each velocity bin ($dN_v$) is then found by
\begin{equation}
dN_v=\tau_v\times\frac{8\pi\nu^2_0}{c^2}\frac{Z}{g_u}\frac{1}{A_{ul}}\exp\left(\frac{E_l}{kT_\textrm{ex}}\right)\left[1-\exp\left(-\frac{h\nu_0}{kT_\textrm{ex}}\right)\right]^{-1} dv\label{eq:dN_v}
,\end{equation}

where $A_{ul}$ is the Einstein coefficient for each transition, $E_l$ is the energy for the lower level of each transition, and $g_u$ the degeneracy of the upper lever of the transition.  $\tau_v$ is derived from the brightness temperature at each velocity bin ($T_{mb}(v)$) by

\begin{equation}
\tau_v=-\ln\left[1-\frac{kT_\mathrm{B}(v)}{\eta h\nu}\left\{\exp\left(\frac{h\nu}{kT_\mathrm{ex}}\right)-1\right\}\right] \label{eq:tau_v}
,\end{equation}

and the partition function Z is
\begin{equation}
Z=\sum_{l=0}^\infty g_l \exp\left(-\frac{E_l}{kT_\textrm{ex}}\right)
.\end{equation}

The main spectroscopic parameters of these equations are shown in Table \ref{tablespecSum}. 
   \begin{table}
      \caption[]{Spectroscopic parameters}
         \label{tablespecSum}
                \begin{center}
        \begin{tabular}{llllll}
            \noalign{\smallskip}
            \hline
            \noalign{\smallskip}
            Species      &  Transition & A$_{ul}$ [s$^{-1}$] & g$_u$  & \\
            \noalign{\smallskip}
            \hline
            \noalign{\smallskip}
        $^{13}$CO    & J=3-2 &  $2.19\times10^{-6}$ & 7 & \\ 
            \noalign{\smallskip}
        $\ci$   & $^3$P$_1$ - $^3$P$_0$ &  $7.98\times10^{-8}$ & 3 & \\                                        \noalign{\smallskip}
        $\cii$  & $^2$P$_{1/2}$ - $^2$P$_{1/2}$         &  $2.29\times10^{-6}$ & 4 & \\ 
            \noalign{\smallskip}
            \hline
            \end{tabular}
                \end{center}
Note: Values from CDMS \citep{muller01}. 
      \end{table}

\begin{figure*}[!hpt]
\includegraphics[angle=0,width=18cm]{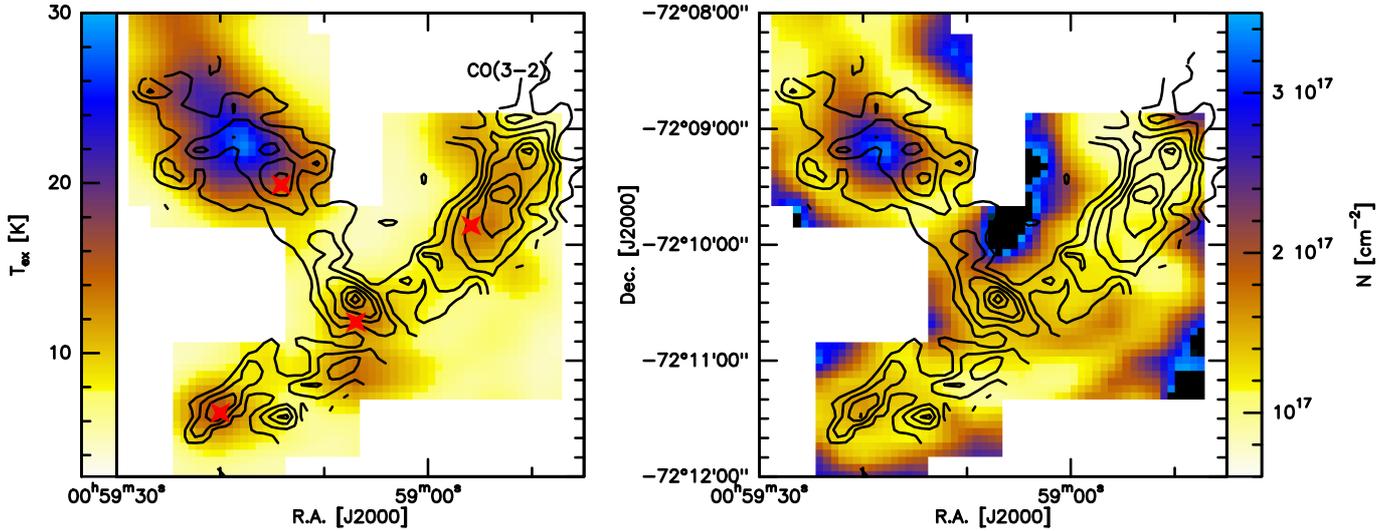}
  \caption{\footnotesize{Maps of the LTE excitation temperature and column density for the CO molecule in N~66, calculated assuming a $\eta$ of 0.1 to account for the filling factors. Left: CO excitation temperature; stars represent the positions that have been used in Table~\ref{tablelte}. Right: CO column densities. In all panels, black contours refer to the $\cii$ integrated intensity as in Fig. \ref{HAlpha}.}}
  \label{N66_CO_TN}
\end{figure*}

We calculated the $^{13}$CO column density and then converted this into the CO column density by multiplying it with the isotopic abundance ratio [$^{12}$C]/[$^{13}$C]=50 (see Sect. 3.2.2).  Total column densities $N$(CO) were then derived by integrating $dN_v$(CO) over the frequency range corresponding to the velocity interval 141--166 $\kms$. The resulting maps of excitation temperature and column density across N~66 are shown in Fig.~\ref{N66_CO_TN}.

%

The carbon excitation temperatures and column densities are found in a similar way from the same basic equations \citep[see][Eqns. (4) and (5)]{okada15}. However, here we are hindered by the fact that there is only a single $\cii$ transition and that out of the two $\ci$ transitions, we have observed only one (the lower frequency J=1-0 transition at 492 GHz). Therefore we cannot use the same approach.

The $T_{ex}$ of $\ci$ and $\cii$ have to be assumed. First, as CO and $\ci$ appear to be closely associated, we assumed identical excitation temperatures for both. Furthermore, we assumed that the $\cii$ emission is in the high-temperature, high-density limit, so that Eq. \ref{eq:dN_v} yields a lower limit to the actual column density.


%
   \begin{table}{}
      \caption[]{Results of the LTE analysis (for details, see Sect. 3.2.1) for different assumed filling factors for the C-bearing species in N~66, N~25, and N~88 }     
         \label{tablelte}
                \begin{center}
        \begin{tabular}{l|c|c|c|c}
            \noalign{\smallskip}
            \hline
            \hline
            \noalign{\smallskip}
Pos. &\multicolumn{4}{c}{Beam-averaged column densities} \\
            & $N(\cii)$ &$N(\ci)$  & $N(CO)$ & $N_H$ \\
&\multicolumn{3}{c}{[10$^{17}$cm$^{-2}$]}&[10$^{22}$cm$^{-2}$] \\
            \noalign{\smallskip}
            \hline
\multicolumn{5}{c}{$\eta$ = 0.1}\\
            \hline
N66 
plume & 0.62 & 0.42 & 0.09 & 1.18 \\
ridge & 0.67 & 0.25 & 0.10 & 1.05 \\
N25 & 0.93 & 0.42 & 0.19 & 1.61 \\
N88  & 1.02 & 0.27 & 0.12 & 1.45 \\
                        \hline
\multicolumn{5}{c}{$\eta$ = 0.2}\\
            \hline
N66 
plume & 0.93 & 0.66 & 0.10 & 1.74\\ 
ridge & 0.99 & 0.38 & 0.11 & 1.54 \\
N25 & 1.34 & 0.59 & 0.20 & 2.20 \\
N88  & 1.33 & 0.38 & 1.30 & 1.88 \\
                        \hline
\multicolumn{5}{c}{$\eta$ = 0.5}\\
            \hline
N66 
plume & 0.87 & 2.03  & 0.15 & 3.10 \\
ridge & 0.93 & 0.98  & 0.15 & 2.20 \\
N25 & 1.25 & 1.42 & 0.28 & 3.05 \\
N88  & 1.20 & 0.95 & 0.19 & 2.40 \\
                        \hline
\multicolumn{5}{c}{$\eta$ = 1.0}\\
            \hline
N66 
plume & 0.85 & 2.60 & 0.22  &5.5 \\
ridge & 0.93 & 2.60 & 0.23 & 3.9 \\
N25 & 1.22 & 3.64 & 0.40 & 5.4 \\
N88  & 1.16 & 2.33 & 0.27 & 3.9 \\
                        \hline
            \end{tabular}
                        \end{center}
     \end{table}

 The CO cloud at the plume tip is relatively highly excited, but the LTE excitation temperatures are low as $\sim$8, 12, 20, and 30 K for filling factors of 1, 0.5, 0.2, and 0.1, respectively.  

Table \ref{tablelte} shows the calculated column densities in the main emission components of the $\cii$ map. For the ridge component we have averaged the three main peaks.
The $\cii$ column densities peak in the ridge, but the plume column density is very similar. CO and $\ci$ column densities peak in N~66C, but the CO clumps in the ridge also stand out. The conversion factor between the T$_{mb}$ emission and the column densities are $5.2\times10^{15}$ and $4.0\times10^{16}$ $cm^{-2}/K$ for $\eta=$1 and 0.1, respectively.

\subsubsection{Results from radiative transfer modeling}

Especially for the CO cloud in the plume tip (N~66C), there is sufficient material available to make a more sophisticated analysis using the publicly available non-LTE radiative transfer code $\it RADEX$. It uses as input the space density and the kinetic temperature of the parent molecular hydrogen gas, and the column density of the species under consideration to calculate as output the line flux density that will be observed. In addition to being one-dimensional, the code assumes an isothermal and homogeneous medium without any {\it \textup{large-scale}} velocity gradients. It is comparable to the LVG-method. We used it to match predicted line fluxes as a function of input parameters to the five observed transitions of $^{12}$CO presented in this paper, and the observed $^{12}$CO/$^{13}$CO isotopic ratios in the $J$=1-0 and $J$=2-1 transitions from the {\it SEST} program \citep[cf. ][]{rubio00,israel03} and the $J$=3-2 data from this paper.  For the occasion, we re-reduced the {\it SEST} data. For the N~66C cloud (plume) we found a reasonably good fit of the model data to the observations for a kinetic temperature $T_{kin}(H_{2}) = 40 K$ and a density $n(H_{2})\,=\,9\times10^{3}$ cm$^{-3}$ (see Table\,\ref{lvgtable}). The fits allow the [$^{12}$CO]/[$^{13}$CO] isotope abundance ratio to lie between 40 and 80, but a ratio of 50 gives - as we assumed above - the best result. In this case, the beam-averaged CO column density is $N(CO)\,\approx\,4\times10^{16}$ cm$^{-2}$ for a filling factor $\eta(CO)\,\approx\,0.4$.  All the data suggest that the $\ci$ emission is associated with the same molecular gas as CO, which means that it is characterized by the same molecular hydrogen kinetic temperature and volume density. In this case, the neutral carbon column density is $N(\ci)\,\approx\,1\times10^{17}\times\eta(\ci)$ cm$^{-2}$. For equal filling factors $\eta$ of CO and $\ci$, both species will therefore have the same beam-averaged column density.

Again assuming a gas phase [C]/[H] abundance of  $0.4\times10^{-5}$ , we find the mass of the molecular cloud associated with the observed $\ci$ and CO emission to be about $M(H_{2}+He)\,=\,8\times10^{3}$ M$_{\odot}$.

We performed the same analysis for the CO in the ridge component. Here, we have only a single observed $^{12}$CO/$^{13}$CO isotope ratio ($J$=3-2), whereas the $^{12}$CO $J$=6-5 transition is poorly observed. Nevertheless, we find a reasonably good fit for $T_{kin}(H_{2})$ = 50 K and a density $n(H_{2}$) = $3\times10^{3}$ cm$^{-3}$, that is, somewhat hotter and less dense than the cloud at the plume tip (see Table \,\ref{lvgtable}). The beam-averaged CO column density is $N(CO)\,=\,2.2\times10^{16}$ cm$^{-2}$ for a smaller beam filling factor $\eta(CO)\,\approx\,0.081$. With the same small beam filling factor, the neutral carbon-averaged column density would be a factor of three lower, that is, $N(\ci)\,\approx\,7\times10^{15}$ cm$^{-2}$.  The molecular gas mass associated with the ridge would be $M(H_{2}+He)\,\approx\,7\times10^{3}$ M$_{\odot}$, slightly lower than that of the plume cloud despite its much larger extent. None of this takes into account molecular gas that is not associated with CO emission, but traced by $\cii$ emission instead. Below we show that this CO-dark contribution is not only significant, but even dominant.

The values derived with the radiative-transfer method are, in general, similar to those obtained with the simpler LTE method. Differences are caused by the fact that radiative-transfer modeling is more realistic than the assumption of LTE, which is  certainly not quite correct. Differences are also caused by the more realistic assumption of identical CO and $\ci$ kinetic temperatures rather than identical excitation temperatures, and by a more realistic estimate of $\cii$ parameters. In the LTE modeling the different carbon phases (CO, $\ci$, and $\cii$) 
were treated as unrelated to one another, whereas in the LVG modeling we treated them as much as possible as a coherent ensemble, resulting in a more constrained solution than obtained with the LTE modeling. For example, the LVG modeling allows us to define the CO filling factor and to constrain the isotopic abundance, both of which had to be assumed in the LTE treatment.

Finally, we note that the temperature and density model results obtained for both the plume and the ridge predict a ratio of 0.75 for the $\ci$ $^3$P$_2$ - $^3$P$_2$/$^3$P$_1$ - $^3$P$_0$ ratio, independent of the actual $\ci$ column density. Determination of this ratio thus provides an independent check on the results derived here.

\subsection{Compact object N88}

\begin{figure*}[!hpt]
         \includegraphics[angle=0,width=18cm]{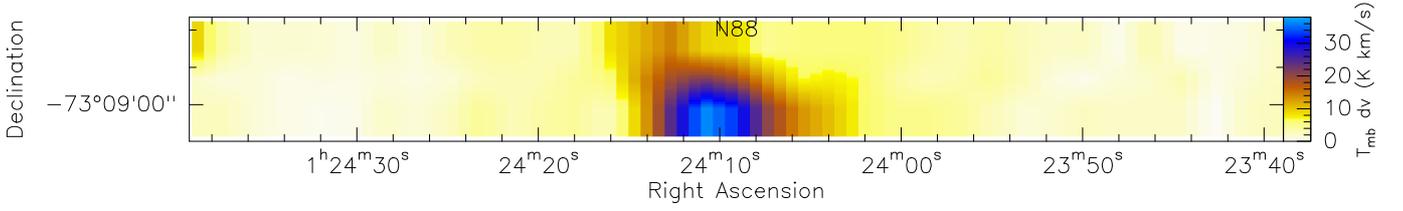}
 \caption{\footnotesize{N~88 total $\cii$ emission strip map obtained with {\it GREAT}, by integration over the velocity range $V_{LSR}$=142.5-157.5 km/s. The $\cii$ emission is mostly limited to a bright peak coincident with the compact $\hii$ region, although it is surrounded by a small region of lower intensity. There is very little $\cii$ emission in the rest of the map.}}
  \label{N88CIIcut}
\end{figure*}

\begin{figure*}[!hpt]
\begin{center}
\vspace{0.25cm}
\includegraphics[angle=0,width=16cm]{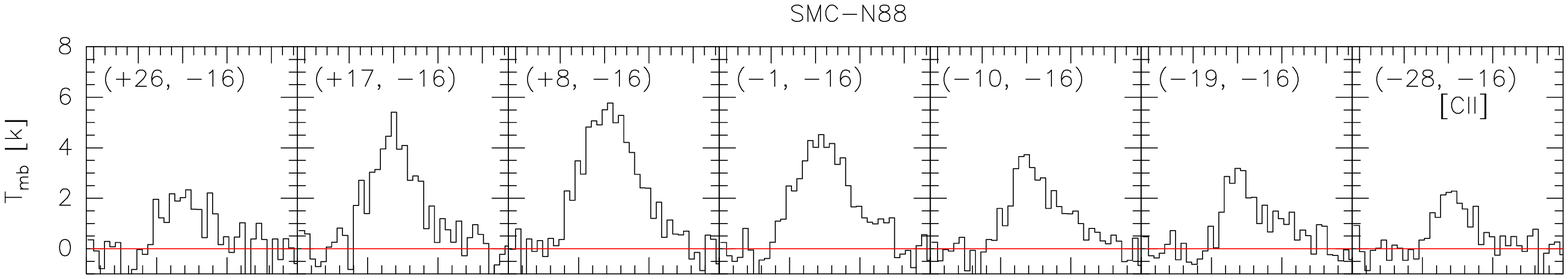}
\vspace{0.25cm}
\includegraphics[angle=0,width=16cm]{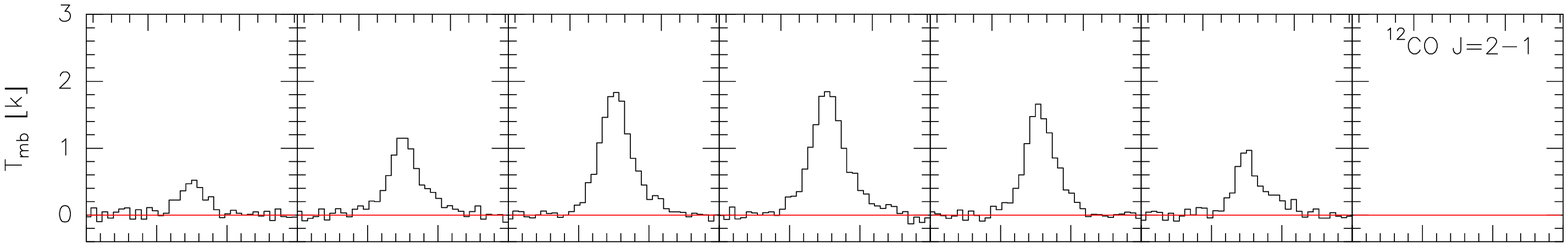}
\vspace{0.25cm}
\includegraphics[angle=0,width=16cm]{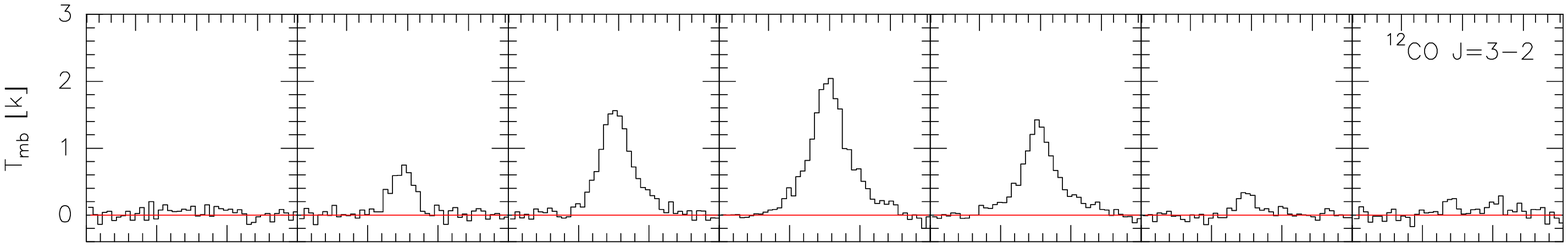}
\vspace{0.3cm}
\includegraphics[angle=0,width=16cm]{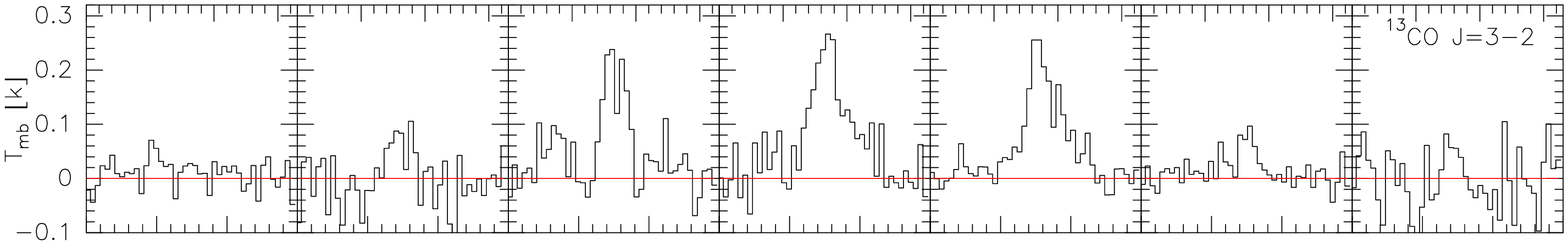}
\vspace{0.25cm}
\includegraphics[angle=0,width=16cm]{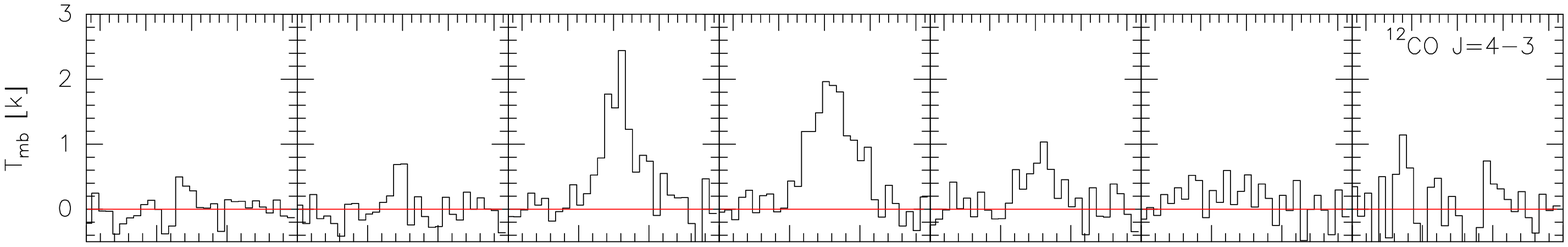}
\vspace{0.25cm}
\includegraphics[angle=0,width=16cm]{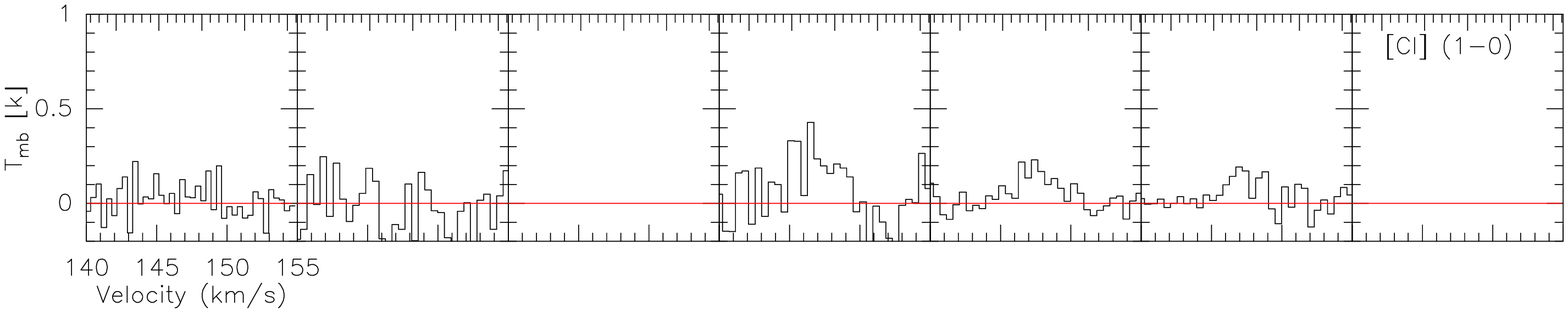}
  \caption{\footnotesize{Observed line profiles along the cut through N~88. Each profile is at the relevant angular resolution as listed in Table \ref{tableObsSum}. The vertical scale is the main beam temperature $T_{mb}$ in Kelvin. The horizontal scale is $V_{LSR}$ in km/s.}}
\end{center}
\label{N88app}
\end{figure*}

The results obtained on N~88 with {\it SOFIA} and {\it APEX} are graphically shown in Figs.\,\ref{N88CIIcut}, \ref{N88app}, and \ref{N88PVmaps}. The $\cii$ map in the former shows a strong resemblance to the $J$=2-1 $^{12}$CO map published by \citet[their Fig. 3]{israel03}, both in extent and morphology. A relatively bright core is surrounded by more diffuse emission extending over about one arcminute ($\approx$ 15 pc), mainly to the west and northeast. We note that the bright and compact $\hii$ region has a diameter of no more than $3''$ \citep[see, e.g., ][]{testor10}, although diffuse H$\alpha$ emission extends over an area of $2.8'\times1.5'$ \citep{davies76}. The compact $\hii$ region is located at $\alpha=01^{h}24^{m}08.0^{s}$, $\delta=-73^{o}09'04''$ ($\Delta\alpha=-2''$, $\Delta\delta=-7''$ in the coordinate system of Fig.\,\ref{N88CIIcut}).  Although the $\cii$ strip map is 4 arcmin long, the central arcmin contains about two thirds of the total measured $\cii$ flux. As the strip map shows, there is widespread diffuse emission in the SMC Wing, with a mean surface brightness of 2.4$\pm$0.4 $\Kkms$ (1.7$\pm$0.3 W m$^{-2}$ sr$^{-1}$).
By averaging profiles over the easternmost and westernmost arcminute (area about 150 pc$^{2}$ each), well away from the star-forming complex, we find emission features about 20-30 $\kms$ wide in velocity. This is much more than the velocity width of the much brighter $\cii$ directly associated with N~88. 

Although coverage and resolution are far from identical, it appears that the integrated flux found here is very similar to that contained in the KAO $\cii$ map published by \cite{israel11}. The $\nii$ map, covering the same area, contains very little emission.  Only at the $\cii$ peak do we detect weak $\nii$ emission at $5\%$ of the $\cii$ ($T_{mb}$) brightness, corresponding to a flux level of about $15\%$.  

The CO radial velocity, $V_{LSR}$ = $147.81\pm0.17$ km s$^{-1}$ ($V_{hel}$ = 151.58 km s$^{-1}$), is remarkably constant over the whole strip.  There is, however, a small but systematic increase in radial velocity from the CO (1-0) to CO (4-3) by 0.38 km s$^{-1}$. The $\ci$ velocity is close to the mean CO velocity, but the $\cii$ falls outside the CO range because it is lower than any of the average CO velocity by $\Delta$v = -0.4 km s$^{-1}$. The $\cii$ emission also shows a hint of a systematic velocity shift with position, characteristic of rotation or outflow (see Fig. A.2).

With increasing $J$-transition, the cloud becomes more centrally concentrated. The FWHM of the velocity-integrated CO emission, corrected for finite beam width by Gaussian subtraction, changes monotonously from $46''$ (13 pc) in the $J$=1-0 transition \citep[data from ][]{israel03} to $11''$ (3 pc) in the J=4-3 transition. This is a much steeper drop than obtained from peak brightness temperatures alone, where the beam-corrected FWHM sizes change from a similar $43''$ (1-0) to about $25''$ (7 pc) with little change between the $J$=4-3 and the $J$=3-2 transitions. These differences represent significant changes in the velocity width of the various profiles. The velocity width of the profiles at the central position increases by a factor of 1.5 from 2.7 km s$^{-1}$ in CO (1-0) to 4.0 km s$^{-1}$ in CO (4-3), while there is almost no corresponding change in the profiles at the neighboring positions.

Although the $\ci$ is rather weak, we find a velocity width of about 3.6 km s$^{-1}$ in the center. The $\cii$ distribution is much wider with FWHM values of $36''$ (velocity-integrated) and $47''$ (peak brightness temperatures). Velocity widths are 4.9 km s$^{-1}$ in the center and drop more slowly than for CO (or for $\ci$).

In all profiles, we also see an asymmetry, with an excess emission wing at the higher (red-shifted) velocities that appears to be most pronounced at the central position, and in the (low optical depth) $^{13}$CO(3-2) and $\cii$ lines, and the (high optical depth) CO (4-3) line.

In the same way as for N~66, we modeled with {\it RADEX} the line ratios available for N~88 and found a fit for $T_{kin}(H_{2})$ = 30 K, $n(H_{2})$ = $3\times10^4$ cm$^{-3}$, once again a $^{12}$CO/$^{13}$CO isotopic ratio of 50, a beam-averaged column density $N(CO)$ = $2.6\times10^{16}$ cm$^{-2}$, and a [C$^{\circ}$]/CO ratio of 0.6 (cf. Table\,\ref{lvgtable}).
\[\]

\subsection{HII region group N~25+N~26}

\begin{figure}[!hpt]
\includegraphics[angle=0,width=8cm]{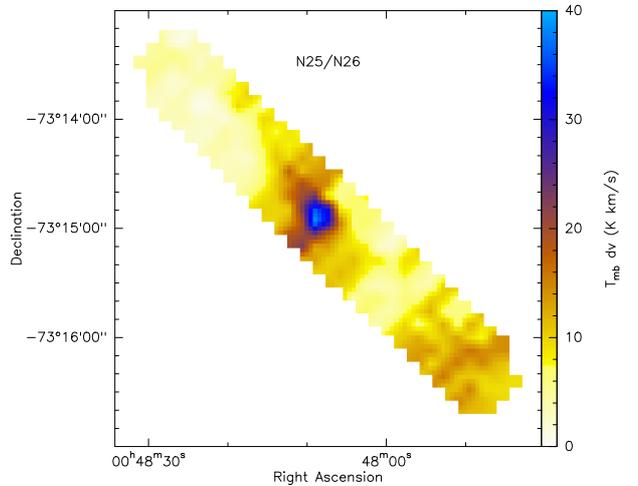}
         \caption{\footnotesize{N~25+N~26 $\cii$ emission strip map obtained with {\it GREAT} by integration over the velocity range $V_{LSR}$ = 110-130 km/s. The bright $\cii$ emission peak at the location of the compact $\hii$ regions is obvious, as is the extended diffuse emission surrounding it. The weak emission in the southwest is associated with the N~22 molecular cloud. }}
  \label{N25CIIcut}
\end{figure}

\begin{figure*}[!hpt]
\begin{center}
\vspace{0.25cm}
\includegraphics[angle=0,width=16cm]{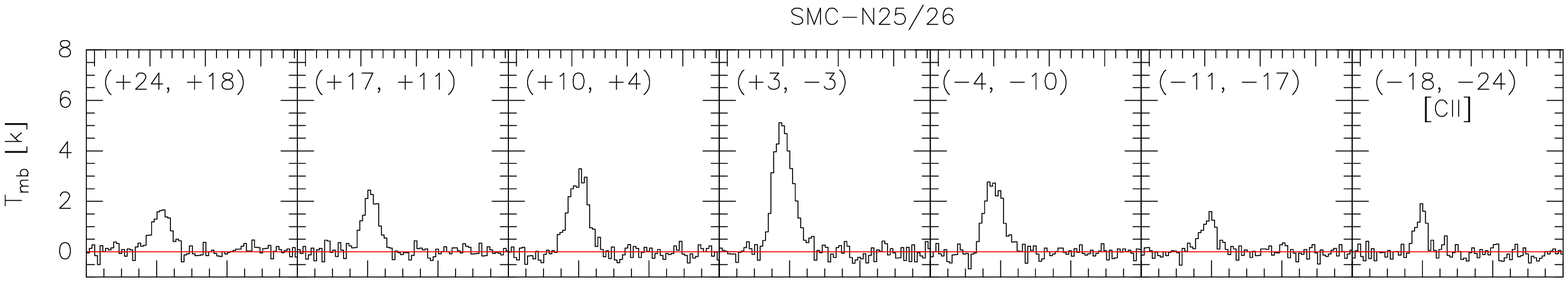}
\vspace{0.25cm}
\includegraphics[angle=0,width=16cm]{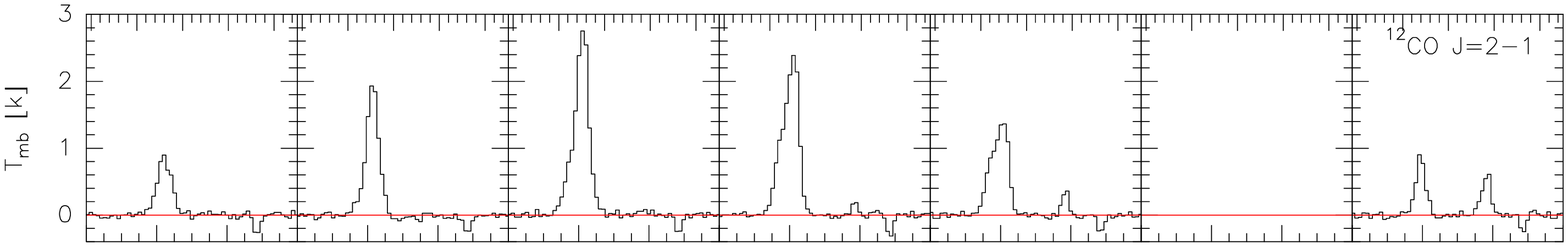}
\vspace{0.25cm}
\includegraphics[angle=0,width=16cm]{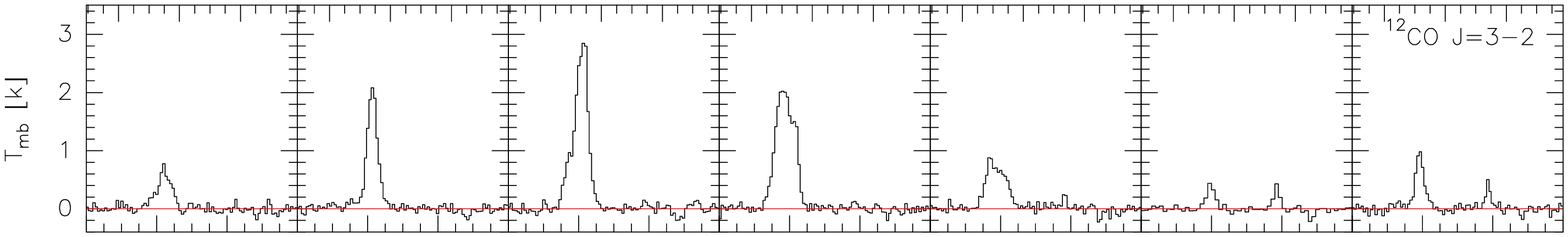}
\vspace{0.25cm}
\includegraphics[angle=0,width=16cm]{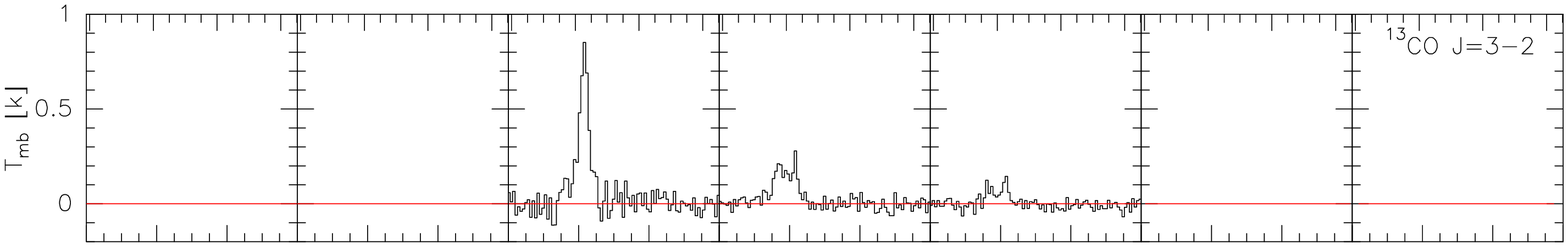}
\vspace{0.25cm}
\includegraphics[angle=0,width=16cm]{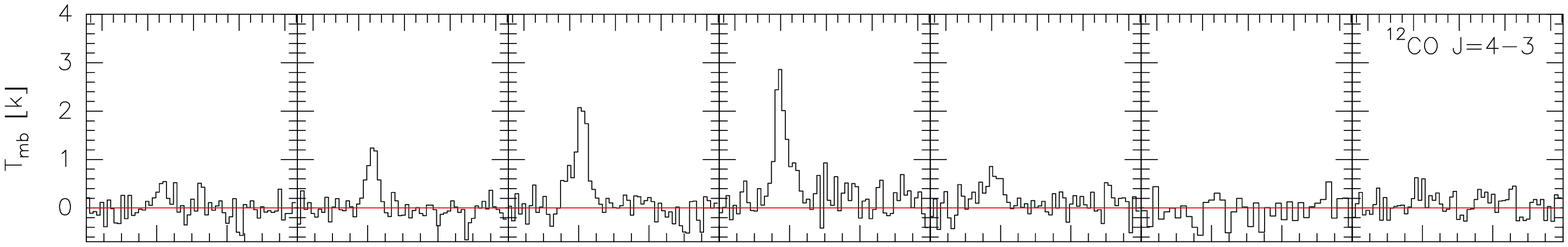}
\vspace{0.25cm}
\includegraphics[angle=0,width=16cm]{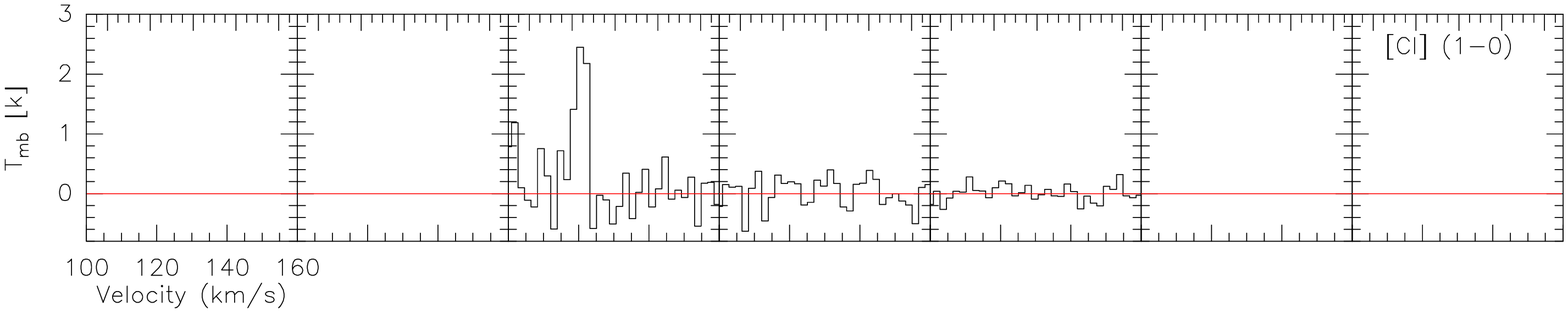}
\caption{\footnotesize{Observed line profiles along the cut through N~25+N~26. Each profile is at the relevant angular resolution as listed in Table\,\ref{tableObsSum}. The vertical scale is main beam temperature $T_{mb}$ in Kelvin. The horizontal scale is $V_{LSR}$ in km/s. }}
\end{center}
\label{N25app}
\end{figure*}

The {\it SOFIA} and {\it APEX} results for N~25+N~26 are displayed in Figs.\,\ref{N25CIIcut}, \ref{N25app}, and \ref{N25PVmaps}. The [C II] map in Fig.\,\ref{N25CIIcut} is overall very similar to the less detailed {\it KAO} $\cii$ map published by Israel $\&$ Maloney (2011, their Fig. 4), in showing a bright, compact source with a relatively bright extension to the south, superposed on dimmer but much more extended emission to the southwest. Again, the difference in coverage and resolution makes a comparison with the KAO map by \cite{israel11} difficult, but it appears that the map shown here recovers between a third and half of the total flux contained in the $3'\times3'$ KAO map. The {\it GREAT} $\cii$ peak coincides with the compact CO source SMC-B2 3, which is part of a larger complex in the southwestern bar of the SMC mapped by \citet[Figs. 3 and 4]{rubio93}. Like the $\cii$ map, the CO map shows extended emission to the west and south and only little extent toward the northeast. The overall distribution of the CO emission and the far-infrared dust emission \citep{bot10} are very similar. The optical images presented by \cite{testor01} and \cite{testor14} show that the main sources of energy heating the compact cloud are the exciting O stars of the diffuse and evolved HII region N~25 (diameter $41''$ or 12 pc) and the compact multiple group of knots N~26 A, B, and C (overall diameter $5''$ or 1.3 pc) that is more dust-rich with $E(B-V)=0.6$. The peak of $\cii$ emission coincides with the compact $\hii$ region located at $\alpha=00^{h}48^{m}08.60^{s}$, $\delta=-73^{o}14'54.7''$ ($\Delta\alpha=+3''$, $\Delta\delta=-6''$ in Fig.\,\ref{N88CIIcut}, as reference for the spectral maps) \citep{testor14}.  The compact ($\approx30''$) and bright $\cii$ cloud shown in Fig.\,\ref{N25CIIcut} is surrounded by a larger region ($90''$) of lower surface brightness. Toward
the southwest, the $\cii$ brightness does not drop below $\int\,T_{mb}$d$V$ = 8-10 $\Kkms$. A region of somewhat higher $\cii$ integrated intensities (13-15 $\Kkms$) similar to that of the N~25+N~26 envelope can be seen in the lower quarter of the {\it GREAT} strip map. This second $\cii$ region is not covered by our CO cuts, but it borders on a region of relatively low CO (1-0) intensity in the maps published by \cite{rubio93}. To the northeast of the compact $\hii$ regions we still find widespread low-surface brightness $\cii$ emission, on average at 2-3 $\Kkms$.

The $\cii$ and $\nii$ maps cover the same area, but we did not detect any emission in the latter.  At the $\cii$ peak, the $\nii$ intensity is below $2\%$ of that of the $\cii$ line ($\leq\,4.4\%$ in flux per beam). The CO line profiles southwest of the N~25+N~26 peak show a second line component at a velocity about 20 km s$^{-1}$ higher ($V_{LSR}\approx139$ km s$^{-1}$).  This velocity range was not included in the maps published by \cite{rubio93}. However, inspection of the original {\it SEST} data reveals that the line is the diffuse outer part of an extended CO cloud peaking at roughly $\Delta\alpha=+30''$, $\Delta\delta=-60''$ , that is the complex consisting of the HII regions N21, N22, and N23, south of N~25+N~26 \citep[cf. ][]{testor14}. The CO emission from this diffuse cloud only has a very weak counterpart in $\cii$ emission.  In the following, we do not consider this component.

%

The radial velocity changes across the cloud from $V_{LSR}$ = $122.2\pm0.1$ km s$^{-1}$ ($V_{hel}$ = 132.7 km s$^{-1}$)\, to $V_{LSR}$ = $119.4\pm0.1$ km s$^{-1}$ in the best-defined $J$=2-1 and $J$=1-0 transitions. This corresponds to a velocity gradient of 2.8 km s$^{-1}$ over $60''$ (17.5 pc). However, the two measured positions closest to the HII regions, (+3, -3) and (-4, -10) have radial velocities lower than expected from this gradient that {\it \textup{decrease}} even more with CO rotational transition: $J$=2-1, $\Delta$V = -0.5 km s$^{-1}$, $J$=3-2, $\Delta$V = -1.8 km s$^{-1}$, and $J$=4-3, $\Delta$V = -2.9 km s$^{-1}$. The quality of the $\ci$ data is insufficient to recognize a pattern in velocities, but for the $\cii$ emission, an even more outspoken pattern is obvious. At the larger position offsets, the [C II] velocity is lower by about 0.5 km s$^{-1}$ (blueshifted) than the CO velocities, whereas it is similar to that of the CO (4-3) velocity at the two positions associated with the HII region.  In the two intermediate offset position, the difference between the CO and $\cii$ velocities is about 1 km s$^{-1}$. The meaning of these patters is not clear. Is the cloud in rotation with a period of about 20 million years? Or are these two partially overlapping clouds with a small velocity difference? In view of the limited resolution of the observations, the crowded nature of the field, and the long lines of sight through this part of the SMC, this is a possibility that cannot be neglected.

The size of the bright cloud is rather the same in the observed transitions. The FWHM of the velocity-integrated CO emission, corrected for finite beam width, is $31\pm3''$ (9 pc) in the $J$=2-1 transition, and $22\pm2''$ (6 pc) in the higher CO transitions and in $\cii$, and these results are indistinguishable from those obtained from peak brightness temperatures alone. The velocity width of the two central CO profiles position is greater than at the more distant offsets, typically 6.5 km s$^{-1}$ versus 3.0-4.5 km s$^{-1}$. In the $\cii$ line, central velocity widths are likewise 6.5 km s$^{-1}$, but with increasing distance to the emission peak, they do not decrease as much as the CO velocity widths. Overall, the $\cii$ emission characteristics more resemble those of the CO(4-3) transition than those of the CO(3-2) and CO(2-1) transitions. This is also the case for the location of the emission peak itself: in most of the observed lines the highest values are obtained at (+10, +4), which is beyond the compact HII regions, but in both $\cii$ and CO (4-3) this is the case at (+3, -3), i.e. $10''$ (or a projected distance of 3 pc) away and very close to the compact $\hii$ regions.

In some profiles, notably those in the direction of the $\hii$ region, we find an asymmetry. The double-peaked line profiles of the (optically thin) $^{13}$CO(3-2) transition suggest that this asymmetry is due to a blending of two different velocity components about 2 km s$^{-1}$ apart.

Modeling with {\it RADEX} of the line ratios of the N~25+N~26 cloud finally yielded a best fit for $T_{kin}(H_{2})$ = 50 K, $n(H_{2})$ = $1\times10^4$ cm$^{-3}$, an isotopic ratio of 50, a beam-averaged column density $N(CO)$ = $3.7\times10^{16}$ cm$^{-2}$, and a [C$^{\circ}$]/CO ratio of 0.5. These values
are similar to those of N~88 (see Table \ref{lvgtable}).

\section{Discussion}

   \begin{table*}{}
      \caption[]{Physical parameters of the observed molecular clouds derived from LVG analysis.  }
         \label{lvgtable}
                \begin{center}
        \begin{tabular}{l|c|c|c|c|c|c|c|c}
            \noalign{\smallskip}
            \hline
            \hline
            \noalign{\smallskip}
Cloud&&&\multicolumn{4}{c}{Beam-averaged column densities} & &Mass\\
Name&T$_{k}$(H$_{2}$) & n(H$_{2}$) & N(CO) & N($\ci$)$^{a}$ & N($\cii$)$^{b}$ & N$_H^{c}$ & $\eta$ & M(H$_{2}$+He)$^{b}$ \\
& [K] & [10$^{4}$cm$^{-3}$] & \multicolumn{3}{c}{[10$^{17}$cm$^{-2}$]} & [10$^{22}$cm$^{-2}$] & & [10$^{4}$ M$_{\odot}$]\\
            \noalign{\smallskip}
            \hline
N66 plume & 40 & 0.9  & 0.40 & 0.40 & 1.82 (0.4-5.3) & 2.62 & 0.41 & 2.4\\ 
N66 ridge & 50 & 0.3  & 0.22 & 0.07$^{d}$& 0.26 (0.1-1.1) & 0.55 & 0.08 & 1.3\\
N25+N26   & 50 & 1.0  & 0.37 & 0.18 & 2.93 (0.9-7.8) & 3.48 & 0.37 & 3.6\\
N88       & 30 & 3.0  & 0.26 & 0.16 & 4.47 (0.5-7.8) & 4.89 & 0.26 & 3.2\\
            \hline
            \end{tabular}
                \end{center}
Notes: Assuming a 27.3$''$ beam. (a) Assuming H$_{2}$ temperature and density as derived for CO. (b) Calculated following Eqn. 1 from \cite{pineda13}; value adopted assumes temperature $T$ and density $n$ of the molecular hydrogen collision partner as given in Cols. 2 and 3. In parentheses we give the range from the minimum $\cii$ column density in the high-$T$, high-$n$ limit, and the column density if the collision partner H$_{2}$ has a density an order of magnitude lower than given in Col. 3. (c) Assuming the nominal $\cii$ column density and a carbon depletion factor $\delta_{c}$ = 0.4. (d) [CI] data missing for some of the peaks.
      \end{table*}

\subsection{Origin of the $\cii$ emission}

Previous studies of Local Group galaxies, in particular the Milky Way and M33, have shown  clear differences between the beam-averaged profiles of the $\cii$, CO, and $\hn$ lines and of the location and extent of $\cii$ with respect to both CO and H I. For instance, \cite{velusamy14}  found from their {\it Herschel-HIFI} survey of 354 Milky Way sight-lines that there is widespread diffuse $\cii$ emission associated with diffuse molecular (H$_{2}$) clouds faint in CO and diffuse H I clouds. About half of the $\cii$ emission detected in the Milky Way is associated with dense molecular clouds traced by CO. Several sight-lines have $\cii$ profiles in which only part of the emission has a CO counterpart. In the same way, many $\hn$ profiles lack a $\cii$ counterpart over much of their velocity range.  \cite{pineda14} concluded that half of the Milky Way $\cii$ luminosity comes from dense PDRs and $\hii$ regions, and a quarter each from  diffuse molecular gas weak in CO and cold $\hn$. However, these results represent very long lines of sight through the Galactic disk, and significant contributions from diffuse and very diffuse gas are therefore not surprising. 

More comparable to the results presented in this paper are the {\it HIFI} $\cii$ maps obtained towards star-forming regions in the dwarf spiral galaxy M~33 by \cite{mookerjea11} and \cite{braine12}. M 33 has a metallicity somewhat lower than the Milky Way in the solar neighborhood \citep[$\sim$factor 2, ][]{magrini10}, and higher than the SMC. In their measurements, much of the $\hn$ gas is also not sampled by either $\cii$ and CO. Although the correspondence between the $\cii$ and CO profiles is much better in the M 33 complexes than in the Milky Way sight-lines, there still is a significant amount of $\cii$ emission in M 33 that does not have a CO counterpart, meaning that it should be associated with diffuse (molecular) gas or, in some cases, $\hii$ regions.

Our study of the $\cii$ emission associated with star-forming regions in  the SMC does not show such effects. On characteristic scales of 10 pc, the emission in the $\cii$ line is clearly and closely related to that in the CO line. They have roughly similar overall extent and peak location. In particular, on the larger scales covered by our maps, we do not find bright $\cii$ emission without a CO counterpart. 

However, on smaller scales of a few parsec, distinctions become apparent. The individual emission peaks of $\cii$ and CO can be offset from each other by a few parsecs.  There is a clear tendency for $\cii$ emission to be relatively weak where CO emission is strong, and strong where CO is weak. In addition, the $\cii$ emission is more widely distributed around the stellar heating sources (i.e., the CO emission is more strongly peaked), and the $\cii$ envelopes are more pronounced than the corresponding CO envelopes. 

The $\ci$ emission sampled tends to follow the distribution of CO, and in particular that of the $^{13}$CO isotope. In addition, $\cii$ and $\ci$ appear to avoid one another, that is, they are somewhat complementary in the maps. The $\cii$ profiles have velocity widths {\it \textup{greater}} than those of CO by up to $50\%$. In contrast, the $\ci$ profiles (and those of $^{13}$CO) have velocity widths {\it \textup{narrower}} than those of CO by about $20\%$.

In the SMC objects studied here, the behavior of $\cii$ with respect to $\ci$ and CO is roughly similar to that seen in the significantly less metal-poor LMC $\hii$ region complex N~159 in the LMC (Okada $\etal$ 2015). For instance, although $\cii$ is also dominant in the N~159, it is so to a lesser degree. $\cii$ and CO also tend to complement each other in N~159. The respective line profiles do not resemble each other nearly as well as in the SMC. A similarity of line profiles in $\cii$ and CO is not found in other recent spectrally resolved extra-Galactic and Galactic studies either, such as in IC342 \citep[][in prep.]{roellig15} and M17 \citep{perez15}.
 
\subsection{Ionized carbon column densities in atomic and molecular hydrogen gas}

The data presented in this paper allow us to make some rough but quantitative estimates for the ionized carbon column densities associated with the various hydrogen gas phases: neutral hydrogen $\hn$, ionized hydrogen $\hii$, and molecular hydrogen $H_{2}$. 

\subsubsection{Ionized carbon in $\hii$ regions}

First, we estimate the possible contribution by $\hii$ zones. From the {\it ATCA} radio maps by \cite{reid06} we find typical thermal fluxes of about 6 mJy per $15''$ beam for the N~66 ridge and plume. This yields an emission measure E.M. ( =$n_{e}^{2}d$) to be $1.6\times10^{4}$ pc cm$^{-6}$ \citep{Mezger67}. Dividing by the electron density $n_{e}\,=\,220$ cm$^{-3}$ \citep{heydari10}, we find an electron column density $N_{e}\,=\,2.2\times10^{20}$ cm$^{-2}$ , which in an $\hii$ region is equal to the hydrogen column density $N_{H}$. With the carbon elemental abundance used before, this turns into a carbon column density $N_{C}$ = $5\times10^{15}-6\times10^{15}$ cm$^{-2}$. This is an upper limit to the column density of C$^{+}$ that can be present in the ionized gas in N~66, because not all carbon in the $\hii$ region is singly ionized. As this upper limit is already one or two magnitudes below the  C$^{+}$ column densities in Tables \,\ref{tablelte} and \ref{lvgtable}, it is clear that the contribution of the N~66 $\hii$ region to the inferred ionized carbon column densities is at most a few per cent - independent of its contribution to the observed $\cii$ emission.  

From the {\it ATCA} radio continuum source catalogs by \cite{filipovic02} we find a thermal flux density of 110 mJy for the $8''$ compact $\hii$ region N~88, with an electron density $n_{e}\,\geq\,3000$ cm$^{-3}$ \citep{kurt99,testor03}. From this we derive a column density upper limit $N_{H}\,\leq\,8\times10^{20}$ cm$^{-2}$, or $N_{C}\,\leq\,2\times10^{16}$ cm$^{-2}$. Beam dilution decreases this even further to $N_{C}\,\leq\,5\times10^{15}$ cm$^{-2}$. For the source complex N~25+N~26 the numbers are less clear, but we estimate in a similar way from the radio flux \citep{filipovic02} and the electron densities \citep[$\approx\,10^{3}$ cm$^{-3}$ -][]{testor14} an emission measure E.M. = $7\times10^{4}$ pc cm$^{-6}$ and column densities $N_{H}\,=\,2\times10^{20}$ cm$^{-2}$ , from which we find $N_{C}\,\leq\,2\times10^{15}$ cm$^{-2}$ once again after taking beam dilution into account.
In either case, the contribution of the $\hii$ region to the (ionized) carbon column density is not more than 2-5 per cent. Given the uncertainties in the analysis of the $\cii$ observations, the contribution of the $\hii$ regions is negligible.

\subsubsection{Ionized carbon in the $\hn$ gas}

Maps of the neutral hydrogen distribution in the SMC based on {\it Parkes} and {\it ATCA} measurements\footnote{The Australia Telescope Compact Array and the Parkes radio telescope are part of the Australia Telescope National Facility which is funded by the Commonwealth of Australia for operation as a National Facility managed by CSIRO. This paper makes use of archived data obtained through the Australia Telescope Online Archive (http://atoa.atnf.csiro.au).}  with a resolution of an arc-minute have been published by \cite{stanimirovic99}. We extracted $\hn$ line profiles at the plume and ridge positions of N~66 and at the positions of N~88 and N~25.   As part of the extraction procedure, the $\hn$ line column densities were automatically determined from the integrated profiles. These are very similar for the positions observed: $N(\hn)$ is about $4.8\times10^{21}$ cm$^{-2}$ for N~88 and the N~66 ridge positions, $3.5\times10^{21}$ cm$^{-2}$ for the N~66 plume, and $1.0\times10^{22}$ cm$^{-2}$ for the N~25 sight-line through the SMC southwest Bar.  In all cases, the $\hn$ line emission covers a much wider velocity range than the CO and $\cii$ lines at the same positions.
In the $\hn$ profiles, only ambiguous peaks are found at the  velocities of the well-defined CO and $\cii$ peaks.

%
\begin{figure}[!hpt]
\begin{center}
\includegraphics[angle=0,width=9cm]{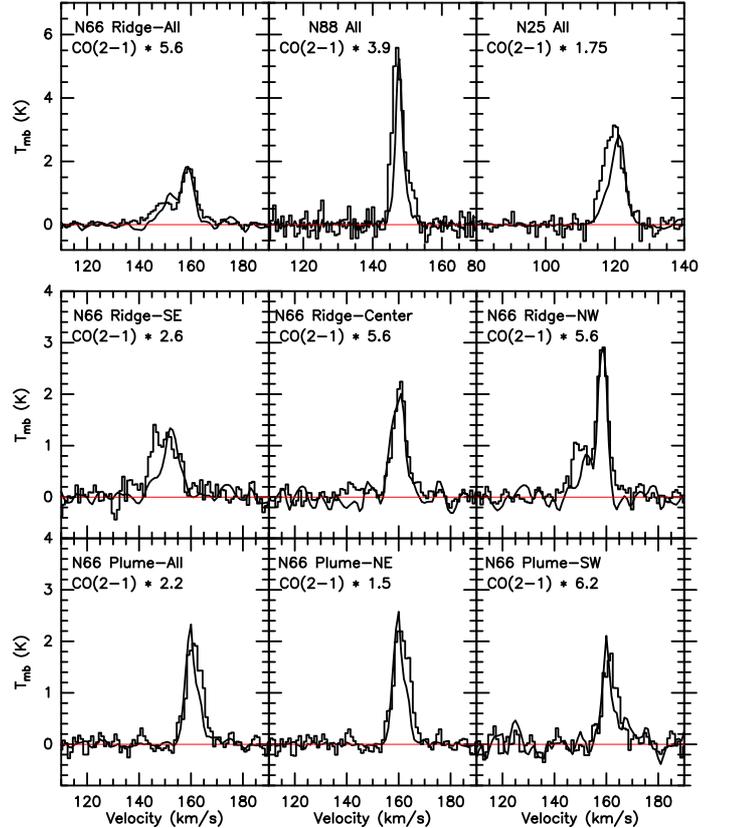}
\end{center}
\caption{\footnotesize{Comparison of observed $\cii$ (histogram) and CO (2-1) (continuous line) profiles from SMC star-forming regions. In each panel the spatially averaged (27.3$''$) profile is shown according to the legend in the upper left corner. The
vertical scale is in main-beam brightness temperature $T_{mb}$ in Kelvin. All CO (2-1) amplitudes have been multiplied by the factors indicated in each panel to match the $\cii$ profile as closely as possible. We note the very close overall similarity of $\cii$ and CO profiles.
}}
  \label{multifig}
\end{figure}

For each of the $\hn$ profiles we determined by scaling which fraction of the integrated $\hn$ flux overlaps with the $\cii$ profiles and thus may be associated with $\cii$ emission.  This fraction is typically about $30\%$, with the exception of the N~66 plume, where it may be as high as $50\%$. From the velocity-integrated $\hn$ flux densities thus found, we calculated the corresponding carbon column densities by applying the elemental carbon abundance [C]/[H] = $2.5\times10^{-4}$ relevant for the SMC and a carbon depletion factor of 0.25 \citep[which assumes that $75\%$ of the carbon in the diffuse neutral gas is locked up in dust grains as in the Milky Way - see][]{jenkins09}. These carbon column densities are (0.5-1.0)$\times10^{16}$ cm$^{-2}$ for N~66, 0.9$\times10^{16}$ cm$^{-2}$ for N~88, and 2$\times10^{16}$ cm$^{-2}$ for N~25+N~26. 
These carbon column densities are lower by (much) more than an order of magnitude than the total carbon column densities in Table 4. 
Because the ionization potentials of hydrogen and carbon are very close, we expect that only a relatively small amount of the carbon will be ionized in the diffuse $\hn$ gas-phase with its low ionization fraction. Assuming an ionized carbon fraction of $25\%$, we have to apply another factor of four. This yields upper limits of about (1-5)$\times10^{15}$ cm$^{-2}$ to $\hn$-related C$^{+}$ column densities in the objects studied here. It is hard to see how the contribution of the $\hn$ gas to the ionized carbon column densities in the regions studied could be more than a few per cent.  This is comparable to what we have just derived for the contamination from $\hii$ regions. In addition, almost all of the $\hn$ gas in the beam will be at  densities very much lower than the ionized carbon critical density of 3000 cm$^{-3}$, so that the $\cii$ emission from such columns will be wholly negligible.  We conclude that the atomic neutral hydrogen gas phase is likewise a negligible contributor to the observed $\cii$ emission. Although the above carbon column densities estimated to be associated with the $\hn$ gas are in fact  similar to the carbon column densities derived from the $\ci$ line observations, a similar argument holds true: at all conceivable $\hn$ temperatures and densities, the $\ci$ emission would be too weak to detect.

\subsubsection{CO-dark molecular gas.}

In the above, we have found that no more than a very small fraction  of both the $\cii$ emission that we have observed from SMC star-forming regions and the C$^{+}$ column densities implied by this emission is associated with either ionized hydrogen or neutral atomic hydrogen gas. Thus, practically all $\cii$ emission should come from relatively dense molecular gas (although we cannot distinguish between GMCs and PDRs). Given this result, we can now interpret the column densities given in Table\,\ref{tablelte}. For small beam filling factors, most of the molecular $H_{2}$ gas in the CO-bright part of the plume would in fact be traced by CO. If larger beam filling factors apply, as appears to be the case, the amount of CO-dark molecular gas, traced by the carbon lines rather than by carbon monoxide, would increase appreciably. In the remainder of the plume, CO-dark gas is prominent for all beam filling factors, increasing from a third to three quarters of all molecular gas as the beam filling improves. In the ridge section with its small beam filling factors, CO-dark gas is suggested to be dominant (60$\%$ to 70$\%$) except in the center, where it would drop to a third of all molecular gas. 

The radiative transfer results in Table\,\ref{lvgtable} adds
to this, although a definitive result cannot be obtained as we cannot independently determine the physical properties of the gas with which the $\cii$ emission is associated. In the preceding we have established that this is mostly H$_{2}$ gas, and we assume for this CO-dark molecular gas the same kinetic temperature and column density as for the gas directly traced by CO. The C$^{+}$ column densities, calculated from Eq. 1  in \cite{pineda13}, show that the CO-dark gas is dominant in all three SMC regions mapped in $\cii$, $\ci,$ and CO line emission. Its nominal fractional abundance range from about $50\%$ in the N~66 ridge to $85\%-90\%$ in the compact objects N~25+N~26 and N~88. CO itself traces between $5\%$ and $40\%$ of all molecular gas. $\ci$ traces even less, although it cannot be neglected with respect to CO. If the gas associated with the $\cii$ emission is both hotter and denser than the gas associated with CO (which we consider unlikely), the amounts of CO-dark gas drop to about a third of all molecular gas in N~66, but are still over half of all the molecular gas in N~25+N~26 and N~88. If the density of the (molecular) gas associated with $\cii$ were an order of magnitude lower than that of the CO-related gas (but at the same temperature), the fraction of CO-dark gas would in all cases be $80\%-95\%$. Thus, we conclude that most of the molecular gas associated with star-forming regions in the SMC does not produce significant CO line emission, but is bright in the emission of the ionized dissociation product C${+}$.

Although the CO and the $\cii$ lines trace different segments of the H$_{2}$ gas, these segments must be well mixed in the observed gas cloud complexes. As already noted, the outlines of the CO and the $\cii$ emission are very similar. Moreover, as shown in Fig.\,\ref{multifig}, in the observed SMC clouds the averaged CO and $\cii$ line profiles are almost identical in shape and similar in width, unlike the profiles in the more metal-rich Milky Way, M33, and LMC.

\subsection{Expanding ring in N~66}

The position-velocity cuts presented in Fig. \ref{fig:N66slice} show a single velocity component for the plume feature, but a more complex velocity structure in the ridge. Along the ridge, the various features outline a clumpy ring in position-velocity space; this 'ring' is seen in both cuts through the ridge. We
note that this ringlike structure is also easily recognized in the channel maps of Fig.\,\ref{specmaps} and in those published by \citet[their Fig. 5]{rubio00}. In addition, the complex network of shocked $H_{2}$ ridges seen by these authors is specifically associated with this CO/$\cii$ structure. We interpret it as representing an expanding ring or cavity seen edge-on. Its angular diameter of 175 arcsec and its total velocity width up to 20 $\kms$ imply a physical linear radius of 25 pc and an expansion velocity of 7 -10 km s$^{-1}$. An upper limit to its expansion age is therefore 3.5 million years. The expanding bubble model by \cite{weaver77} suggests an actual age of about two million years, taking deceleration into account. These times are very close to the age of three million years estimated by \cite{evans06} for the star cluster NGC~346 on which it is centered. The same model also predicts a line-of-sight column density $N_{H}\,=2.3\times10^{23}$ cm$^{-2}$, which is consistent with the value $N_{H}(beam-averaged)/\eta\,=\,0.7\times10^{23}$ cm$^{-2}$ from Table 4 given the range of uncertainty in that number. As radius, age, and expansion velocity pose only modest energy requirements, it is likely that the expansion is driven by the star cluster NGC~346. 

\cite{contursi00} suggested that NGC~346 creates an interstellar radiation field intensity at the UV wavelength of 160 nm in the center of the ring that is almost an order of magnitude higher than that in the solar neighborhood, dropping by a factor of two to four at the ring edges. At the position of the CO cloud in the plume (their peak G), the UV radiation field does not appear to differ much from the value in the solar neighborhood.  

We thus consider the ridge-like structure and its extension toward the southwest to be the main remaining part of the GMC complex from which the  star cluster NGC~346 and its precursor stars originated. This cloud complex is in an advanced stage of being consumed and destroyed by the stars that it generated. 

The CO plume is a smaller remnant cloud that is likewise being eroded by the radiation from the newly formed stars. However, at its greater distance to the NGC~346 cluster, the impinging radiation field is much weaker and the erosion process correspondingly slower. The molecular cloud at the tip of the plume is only beginning to be heated, and CO is still an abundant and good tracer of the H$_{2}$ present. In contrast, at the tip of the plume and closer to ionizing stars, the CO has disappeared and the bright remaining H$_{2}$ is traced by $\cii$ emission instead. In addition, part of the $\cii$ emission this close to NGC~346 may also be associated with the densest parts of the $\hii$ region.

\section{Conclusion}

The main conclusion from this work is that the bulk of the mass of the  molecular gas associated with the SMC star-forming regions is traced by $\cii$,  not by CO. Our best estimates for the percentage of all molecular gas traced by $\cii$ vary from at least $50\%$ in N~66 to more than $85\%$ in the compact $\hii$ regions N~25+N~26 and N~88. Only if the ionized carbon were to be both very dense and very hot might these percentages be lowered somewhat (to about $35\%$ for N~66 and $55\%$ for the compact objects). Conversely, they would be close to $100\%$ if the $\cii$-emitting gas were to be colder and less dense. We may compare this situation to that found in the Milky Way. The {\it Herschel HIFI} $\cii$ survey of the Galactic plane shows that on average a quarter of the $\cii$ luminosity is associated with CO-dark $\hii$, as opposed to a third of the luminosity being associated with PDRs \citep{velusamy10,pineda14}. In addition, the former found that in more than a third of the clouds surveyed, the amount of molecular gas traced by $\cii$ exceeds that traced by CO. Even more relevant is the conclusion by \cite{pineda13} that the fraction of $\hii$ in the Milky Way that is CO-dark (i.e., primarily traced by $\cii$) ranges from about $20\%$ in the metal-rich inner Galaxy to $80\%$ in the metal-poor outer Galaxy. Thus, the high fractions found here to apply to the observed SMC gas complexes reinforce the implicit conclusion that $\cii$ should be the preferred molecular gas mass tracer in metal-poor environments, unlike CO, which it far outperforms in this respect. However, the rough similarity of the extent of CO, $\ci,$ and bright $\cii$, as well as the general similarity of profile widths, suggests that the distribution of the three species is more closely entwined in the SMC star-forming complexes than in their counterparts in the Milky Way and the LMC.

The origin of the $\cii$ luminosity is less straightforward. The high surface-brightness $\cii$ emission originates almost all from regions dominated by molecular H$_{2}$ gas, with minor contributions from ionized carbon in $\hii$ regions, and perhaps some from $\hn$ directly connected to these. The N~25+N~26 and N~88 strip maps reveal, in addition to the bright $\cii$ associated with the discrete $\hii$ regions, a significant amount of very extended, low surface-brightness $\cii$ emission that was also hinted at by the low-resolution KAO~$\cii$ maps published by \cite{israel11}. Because much of the velocity range covered by the $\hn$ profiles in the same direction shows no discernible $\cii$ emission, it is most likely that the diffuse extended $\cii$ is associated with diffuse molecular gas of relatively low density.

From the above, it is clear that studies of regions at substantially lower metallicities than those in the Milky Way are essential to unravel the physical condition of the molecular ISM as a function of ambient conditions. In this sense, the SMC regions studied here have provided a first look at conditions so extreme that the CO molecule, ubiquitous in the Milky Way, hardly plays a role. Indeed, the SMC provides about the lowest metallicity that can fruitfully be studied: at even lower values, all tracers of molecular hydrogen disappear. The intriguing differences between the regions discussed here and those in the LMC studied  by \cite{okada15} provide further motivation to continue this investigation.

\begin{acknowledgements}
Based [in part] on observations made with the NASA/DLR Stratospheric Observatory for Infrared Astronomy (SOFIA). SOFIA is jointly operated by the Universities Space Research Association, Inc. (USRA), under NASA contract NAS2-97001, and the Deutsches SOFIA Institut (DSI) under DLR contract 50 OK 0901 to the University of Stuttgart.
\end{acknowledgements}

\bibliographystyle{aa}
\bibliography{SMC_GREAT}

\newpage

\begin{appendix}
\section{APPENDIX}
%
\begin{figure}[!hpt]
\begin{center}
\includegraphics[angle=0,width=7cm]{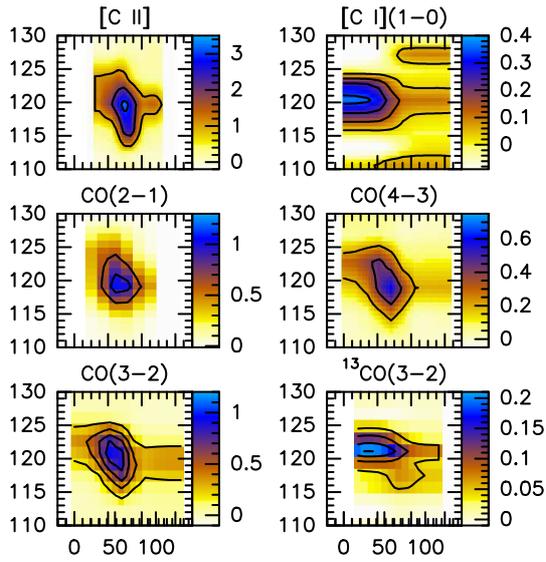}
\end{center}
\caption{\footnotesize{N~25+N~26 position-velocity diagrams in all observed lines. $\ci$ and $^{13}$CO show similar distributions. The CO lines suggest rotation or outflow.}}
  \label{N25PVmaps}
\end{figure}

\begin{figure}[!hpt]
\begin{center}
\includegraphics[angle=0,width=7cm]{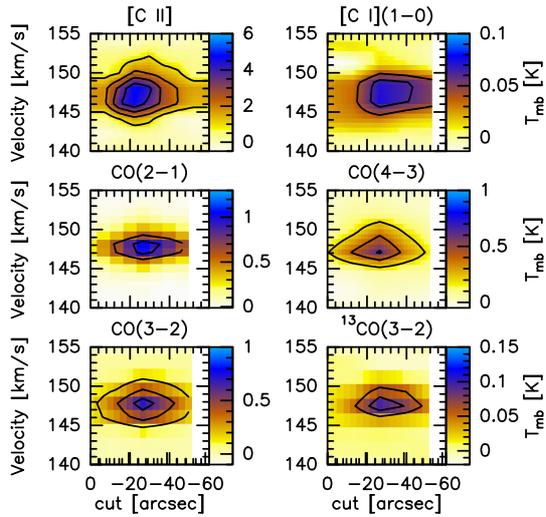}
\end{center}
 \caption{\footnotesize{N~88 position-velocity diagrams in all observed CO, $\ci,$ and $\cii$ lines. The $\cii$ and $\ci$ emission peaks are on opposite sides of the CO peak. The CO emission is more compact than the $\cii$ emission.
 }}
  \label{N88PVmaps}
\end{figure}

\begin{figure}[!hpt]
\begin{center}
\includegraphics[angle=0,width=8cm]{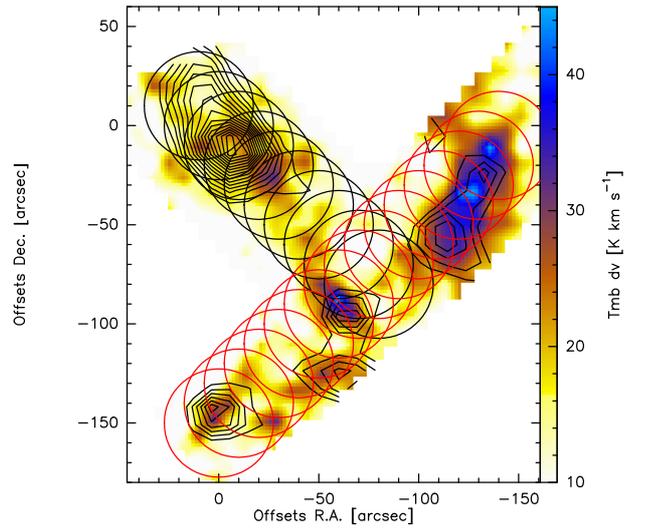}
  \caption{\footnotesize{Beam positions corresponding to the N~66 line profiles shown in Figs.\,\ref{fig:N66pluapp} and \ref{fig:N66ridapp}. The cuts start from the plume to the ridge (north to south) for the plume component and southwest to northeast for the ridge component. All the maps have been convolved to the same 27.3$''$ beam size, and we extracted one beam per position.}}
\end{center}
\label{N66stripori}
\end{figure}

\end{appendix}

\end{document}